\documentclass[prb,preprint,superscriptaddress,showpacs]{revtex4}
\usepackage{amsmath}
\usepackage{amssymb}
\usepackage[]{graphicx}
\usepackage[]{epsfig}
\usepackage[dvipdfm]{hyperref} 
\usepackage{epstopdf}
\def\[[{\left[}
\def\]]{\right]}
\def \intk{\int {d^dk \over (2 \pi)^3}}
\newcommand {\epp} {\epsilon^{\prime}}

\newcommand {\cto} {\right)}
\newcommand {\ato} {\left(}

\newcommand {\be} {\begin{equation}}
\newcommand {\bea} {\begin{eqnarray} \nonumber }
\newcommand {\ee} {\end{equation}}
\newcommand {\eea} {\end{eqnarray}}
 \newcommand {\eps} {\epsilon}
 \newcommand {\si} {\sigma}

 \newcommand {\al} {\alpha}
 
 \newcommand {\N} {{\cal N}}

\newcommand {\lan} {\langle}
\newcommand {\ran} {\rangle}

\newcommand {\cN}  {{\cal N}}

\newcommand {\Tr} {\mbox{Tr}}
\newcommand {\for} {\ \ \ \mbox{for}\ \ }
\newcommand {\sign} {\mbox{sign}}

\def \(  {\left(}
\def \)  {\right(}

\def \form#1 {eq. (\ref{#1}) }
\def \parziale#1#2  {{\partial {#1} \over \partial {#2}}}

 \def \(({\left(}
 \def \)){\right)}

\def \b{\beta}

\def \nn{\nonumber}
\def \beqna{\begin{eqnarray}}
\def \eeqna{\end{eqnarray}}
\def  \beq{\begin{equation}}
\def  \eeq{\end{equation}}

\def \ln{{\rm ln}}

\def \b{\beta}

\newcommand {\sig} {{\bf \sigma}}

\def \ln{{\rm ln}}
\def \ra{\rangle}
\def \ab2{\alpha\beta^2}
\def \la{\langle}

\begin{document}
\title{Glasses and replicas \\
}
\author{Marc M\'ezard$^1$ and
Giorgio Parisi$^2$}
\affiliation{
${}^1$CNRS, Laboratoire de Physique Th\'eorique et Mod\`eles Statistiques,
Universit\'e Paris-Sud, B\^{a}t 100, 91405 Orsay Cedex\\
${}^2$
Dipartimento di Fisica,  SMC and Udrm1 of INFM, INFN,
Universit\`a di Roma ``La Sapienza'',
Piazzale Aldo Moro 2,
I-00185 Rome (Italy)
}
\date{\today}

\begin{abstract} We review the approach to glasses based on the replica formalism.  
The replica approach presented here is a first principle's approach which aims at deriving the main glass properties 
from the microscopic Hamiltonian. In contrast to the old use of replicas in the theory of disordered systems, this replica approach
applies also to systems without quenched disorder (in this sense, replicas have nothing to do with computing the average of  a logarithm of the partition function). 
It  has the advantage of 
describing in an unified setting both the behaviour near the dynamic transition (mode coupling transition) and the behaviour near the 
equilibrium `transition' (Kauzmann transition) that is present in fragile glasses.  The replica method may be used to solve 
simple mean field models, providing explicit examples of systems that may be studied analytically in great details and 
behave similarly to the experiments.  Finally, using the replica formalism and some well adapted approximation schemes, it is possible to do explicit 
 analytic computations of the properties of realistic models of glasses. The results of these first-principle computations are in reasonable agreement with numerical 
simulations.
	\end{abstract}
\maketitle 
	\newpage
	
	\tableofcontents 
\section{Introduction}
\subsection{General considerations} 
In recent years the replica formalism has been brought to bear in the study of glasses.  This may seem paradoxical as the replica method was developed in the seventies in order to study systems with quenched disorder\cite{EA,SK,mpv}, while there is no such quenched disorder in glasses. It turns out that the replica method is actually much more general than is usually thought, and can be used to study systems, like glasses, where disorder is
'auto-induced', in the sense that there exist many `random' equilibrium configurations \cite{MPR,BM94,MONA}. In a nutshell, one can summarize the replica approach as follows: in a glass, it is extremely difficult to describe each equilibrium state, and therefore, we have a priori no theoretical tool which allows us to polarize the system in one of its equilibrium states. However, the system itself 'knows' about these equilibrium states, and therefore one can use some other copy ('replica') of the system as an external field which helps to polarize the system. This initial idea needs to be refined and expanded, in particular as one wants to find out an external field which does not modify too strongly the Boltzmann measure. But it carries the main idea: replicas have nothing to do with computing the logarithm of a partition function! The discovery that replicas can be used in systems where there is no randomness in the Hamiltonian has been a key step which is at the heart of recent progress in the theoretical study of glasses.

 There are many indications that, if we could follow the evolution of a glass at a microscopical level, we would discover that at low temperatures the glass freezes in an equilibrium (or quasi equilibrium) configuration that is highly non-unique.  This essential non-uniqueness of the ground state is present in many others systems where the energy landscape is highly corrugated: e.g. it is widely believed to be present in spin glasses, i.e. magnetic systems with a quenched randomness due to a random mixture of
 ferromagnetic and antiferromagnetic interactions \cite{EA,mpv,BOOK,CINQUE}.  This property is responsible for the peculiar behaviour of glassy systems and, at the same time, it makes the theoretical study of these systems very hard.  The main ingredient of the approach is therefore the statement that glasses may freeze in many microscopically different configurations when we decrease the temperature.  This statement is common to many other approaches \cite{AdGibbs,St}, however the replica approach gives us a panoply of sophisticated physical and mathematical tools that strongly increase our ability to describe, study and compute analytically the properties of glasses.

These replica-based tools have been used in two types of analytic computations. First of all, they allow to compute analytically in an exact and detailed way the properties of some mean-field models for the glass 
transition.  Although theses toy models are somewhat far from reality (the range of the forces is infinite), they display a very 
rich behaviour\cite{KTW}: for example there exist soluble mean field models without quenched disorder  where there is 
an equilibrium glass-liquid transition (Kauzmann transition \cite{kauzman}), a dynamical transition\cite{CuKu,FM,pspin} (mode coupling 
transition(\cite{MCT}) and, at 
a higher temperature, a liquid-crystal transition that cannot be seen in the dynamic of the system (starting from the 
liquid phase) unless we cool the system extremely slowly\cite{MPR,BM94,FMRWZ,BiMez}.  The existence of these soluble models is very precious to us; 
they provide a testing ground of new physical ideas, concepts, approximation schemes that are eventually used in more 
realistic cases.
On the other hand, the replica approach  can also be used to obtain some quantitative results on some realistic models of glasses, starting from a microscopic description in terms of their hamiltonian. These results can be confronted to those of numerical simulations.

The aim of this review is to present an introduction to the replica approach to glasses, and to stress the underlying physical ideas.  The amount of work that has been done in the field is extremely large, and here we will consider only a few most important aspects. Some complementary references to can be found for instance in \cite{GPHouches02,BCKM,Mgold,MeAspen}.

\subsection{Glassiness, metastability and hysteresis}
\label{sect:gmh}
A key feature of glasses is the presence of 
metastability, however it is fundamentally different from the usual case of metastability that we know in non-disordered materials (which can be summarized by: ``a piece of glass is very different from a diamond'').

The `usual'  case of metastability is a system that undergoes a first order phase transition when we change a parameter. When the first order transition happens by changing the temperature, if we cool the systems sufficiently slowly, the high temperature phase survives also below the critical temperature, until the spinodal temperature is reached.

A familiar example is the ferromagnetic Ising model, where the control parameter is the magnetic field $h$ (an other example would be the solid-liquid transition where the control parameter could be the temperature or the pressure). We present this discussion using the magnetic field as control parameter because many of the ideas that we present originate in the study of spin glasses, and in this field there are extensive theoretical analyses, numerical simulations and experiments. At low temperature the equilibrium magnetization $m(h)$ is given by $m(h)=m_{s}\ \sign(h)+O(h)$ for small $h$ ($m_{s}$ being the spontaneous magnetization): the magnetization changes discontinuously at $h=0$ in the low temperature phase where $m_{s} \ne 0$.

Let us consider a system that evolves with some kind of local dynamics.  If we slowly change the magnetic field from 
positive to negative $h$, we enter in a metastable region where the magnetization is positive, and the magnetic field is 
negative.  The system remains in this metastable state a quite large time, given by $\tau(h) \propto \exp( 
A/|h|^{\alpha})$, where $\alpha=d-1$ \cite{METAISI} for an homogeneous system without impurities.  When the observation time is of order of $\tau(h)$ the system 
suddenly jumps into the stable state.  This phenomenon is quite common: generally speaking we always enter into a 
metastable state when we cross a first order phase transition by changing some parameters and the system remains in the {\em wrong} phase for a long time. The time the system remains in the wrong phase strongly depends on the microscopic details and on the presence of impurities that may decrease the height of the barriers and trigger the phase transition,

Let us study further this 'usual' type of metastability in first order transitions \cite{PARISISTAT}.  Starting from  state where $m>0$ at $h=0$, if  we 
add a  {\em positive} magnetic field $h$ at time 0, the linear response susceptibility is  equal to
\begin{equation}
\chi_{LR}= {\lim_{t \to \infty}} \frac{\partial}{\partial h} m(t,h),
\end{equation}
$ m(t,h)$ being the magnetization at time $t$. By linear response theory we find:
\begin{equation}
\beta^{-1} \chi_{LR}=\lim_{h\to 0^+}\sum_{i}\lan \si(i)\si(0)\ran^{c} \equiv 
\lim_{h\to 0^+}\sum_{i}(\lan \si(i)\si(0)\ran-\lan \si(i) \ran \lan \si(0)\ran).
\end{equation}
The linear response susceptibility is {\it not equal} to 
the equilibrium susceptibility that at $h$ exactly equal to zero  is infinite:
\begin{equation}
\chi_{eq}= {\partial\over \partial h} {\lim_{t \to \infty}}  m(t,h){\biggr |}_{h=0} m_{s}
= {\partial\over \partial h}\ \sign(h){\biggr |}_{h=0}=\infty.
\end{equation}
The difference between the two susceptibilities exists only at $h=0$:
\be
\chi_{eq}(h)=\chi_{LR}(h)+ m_{s}\delta(h) \ .
\ee

The introduction of the two susceptibilities adds nothing into the discussion of {\em standard} metastability. We claim that in glassy systems the metastability  is of  a different nature and here the study of the two susceptibilities, $ \chi_{eq}$ and $ \chi_{LR} $  gives  important information.  

First of all, the difference between the two susceptibilities occurs in a whole region of parameter space (the glass phase), not just at a transition. If we consider the case of  spin glasses, it is well known that (at lest in the mean field approximation)  there is an open region in the space of 
parameters, 
where, if we change a parameter of the system (e.g. the magnetic field $h$) by an amount $\Delta h$, we have that 
$\chi_{LR}\ne \chi_{eq}$.  This region, the glass phase, is characterized by $|h|<h_{c}(T)$\cite{AT}.
The function $h_{c}(T)$ increases when we decrease the temperature; $h_{c}(T)$ vanishes at the critical point.

In this region
\bea
\Delta m(t) =\chi_{LR} \Delta h & \for &  1 \ll t \ll \tau(\Delta h),\\
\Delta m(t) =\chi_{eq} \Delta h & \for &  \tau(\Delta h) \ll t,
\eea
where  $\tau(\Delta h)$ may have a power like behaviour
(e.g. $\tau(\Delta h) \propto |\Delta h|^{-4}$).

If we define the \emph{irreversible} susceptibility by 
\be
\chi_{eq}=\chi_{LR}+\chi_{irr} \ ,
\ee
the glassy phase is characterized by a non-zero value of $\chi_{irr}$ \cite{mpv}.  If we observe the system for a time less 
than $\tau(\Delta h)$, its behaviour at a given point of the parameter space depends on the previous history 
of the system, and strong hysteresis effects are present. Note that, in our terminology, hysteresis 
and history dependence do not necessarily imply glassiness.  Hysteresis may be present if the time scale for approaching
equilibrium is very large (larger than the experimental time), but \emph{finite}, has it usually happens at a first order transition.  Glassiness implies an equilibration 
time that is \textit{arbitrarily large}, meaning that it diverges when the system size goes to infinity.  In other words hysteresis can be explained in terms of \emph{finite} free energy barriers 
that may involve only a few degrees of freedom. Glassiness implies the existence of \emph{arbitrarily large} barriers 
that may arise only as a \emph{collective} effect of many degrees of freedom: it can exist only when correlations extend to arbitrary large distances. In the case of hysteresis the barrier are local and may be overcome by the presence of impurities, in the case of  glassiness there must exist  a divergent correlation length.

The physical origine of $\chi_{irr}$ is clear. When we increase the magnetic field, the states with higher magnetization 
become more likely than the states with lower magnetization: this effect contributes to the increase in the 
magnetization.  However the time needed for a global rearrangement of the field is very high (it is strictly infinite in the 
infinite volume limit and for infinitesimal variations of the magnetic fields where non-linear effects are neglected): 
consequently  the time scales 
relevant for $\chi_{LR}$ and $\chi_{eq}$ are widely separated.

\begin{figure}
    \centering
  \includegraphics[width=.6\textwidth]{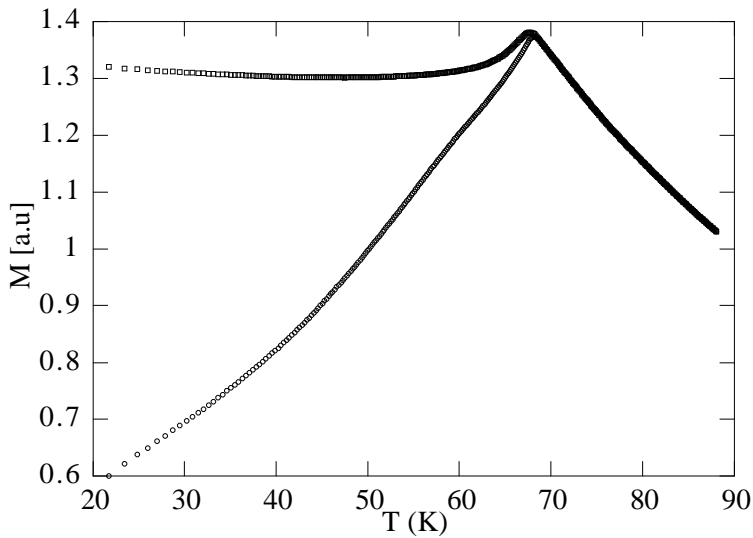} 
\caption{The experimental results for the FC (field cooled) and the ZFC (zero field cooled) magnetisation (higher and 
lower curve respectively) vs.  temperature in a spin glass sample ($Cu_{87}Mn_{13.5}$) for a very small value of the 
magnetic field
$H$ =1 Oe (taken from \cite{EXP}).  For a such a 
 low field,  non-linear effects can  be neglected  and the magnetization is proportional to the susceptibility.}
\label{2CHI}
\end{figure}

The two susceptibilities have been measured  experimentally in spin glasses as follows.
\begin{itemize}
    \item The first susceptibly ($\chi_{LR}$) is measured by adding a very small magnetic field at low temperatures.  This extra field should be small enough in 
order to neglect non-linear effects.  In this situation, when we change the magnetic field, the 
system remains inside a given state and it is not forced to jump from a state to an other state and 
we measure the ZFC (zero field cooled) susceptibility, that corresponds to $\chi_{LR}$.  
 \item
The second susceptibility ($\chi_{eq}$) can be approximately measured by  cooling the system in presence of a small magnetic
field, and comparing the observed magnetization to the one measured without this small magnetic field. In this case the system has the ability to chose the state that is most appropriate in presence of the applied 
field.  This susceptibility, the so called FC (field cooled) susceptibility is nearly independent from the temperature 
(and from the cooling rate - the quasi-independence of the field cooled magnetization on the cooling rate confirms that the field cooled magnetization is near to the equilibrium one-)
and corresponds to $\chi_{eq}$.
\end{itemize}

Therefore one can identify $\chi_{LR}$ and $\chi_{eq}$ with the ZFC susceptibility and with the FC susceptibility 
respectively.  The experimental plot of the two susceptibilities is shown in fig.  (\ref{2CHI}).  They are clearly 
equal 
in the high temperature phase while they differ in the low temperature phase. 
Similar history dependent effects are quite common in structural glasses.

Another characteristic aspect of  glassy systems is the aging of their response functions\cite{POLI,B,BCKM}.  Let us consider an aging experiment 
where the system is cooled and brought to its glass phase at time 0.  

The response function $R(t,t_{w})$ is the variation of an observable at time 
$t_w+t$ in response to a perturbation of the Hamiltonian at a previous time $t_w$. For instance in spin systems the perturbation could be a change in the magnetic field, and the observable could be the magnetization.
 Aging means that the function $R(t,t_{w})$ is not a 
constant in the region where $t$ and $t_{w}$ are \emph{both} large:
\bea
R(t,t_{w})= R_{S} \for t\ll t_{w} \\
R(t,t_{w})= R_{E} \for t \gg t_{w}
\eea
By definition  $R_{S}=\chi_{LR}$ and the identification of $R_{E}$ with $\chi_{eq}$ follows from general 
arguments.

The existence of the glass phase is signalled by  $R_{S}\ne R_{E}$.
This is what is experimentally seen in many experiments done by humans using  values of $t_{w},t$ that are somewhat
shorter than  their life time. The glass phase can be defined experimentally  by the  fact that $R_{S}\ne R_{E}$  on time scales smaller than 10 hours for instance.  These times are much larger (15 or 20 order of magnitude) 
than the microscopic time and the aging effect is quite non-trivial.  Many people do believe that, at low enough temperatures,
 the aging in the response, i.e. $R_{S}\ne R_{E}$, should survive on much larger time scales. A mathematical definition of the glass phase could 
 be that, for an infinitely large system, aging survives in the  limit where $t$ and $t_w$ go to infinity. 

Of course it might be that, in some systems, the glass phase does not exist in the mathematical sense, and the observed aging in experiments is due to 
the fact that they are performed on too short 
times. In this case  aging would disappear in the limits $t,t_{w}\to \infty$.  Knowing for which system, and at what temperatures, the glass phase strictly exists is certainly an important fundamental issue. We will not discuss it here, but rather we will study the physics of glasses starting from the hypothesis that the glass phase exists. Independently of whether the glass phase disappears on infinite time scales, this should be a good starting point to describe the experiments performed on human time scales, where aging is seen. This is the same argument that we would use if we want to study some physical properties (like e.g; the spectrum of phonons) in diamond: it is certainly better to start from the hypothesis that diamond exists, and forget about its actual finite lifetime.

Here we want to focus on the first-principle theoretical study of glasses, starting from some microscopic Hamiltonian. The aim of such studies is to get a theoretical understanding of these effects and to obtain both  qualitative and quantitative predictions. The replica 
formalism is a very efficient way of addressing these issues, as it allows for a simple thermodynamic description of systems
where $\chi_{irr}\ne 0 $. It has the advantage of being very compact and  allowing for detailed explicit computations. On the other hand,
it is very important to always keep in mind the physical meaning of the computation that are being done: the dictionary between the replica computations and the actual physics is non-trivial, and one should always keep it at hand.
   In the glass phase, the difference between the two susceptibilities is in one-to-one correspondence with a phase transition in the replica formalism known as {\bf replica symmetry breaking} and, to the best of our knowledge, it can be explained only in this framework
(or the equivalent framework of the cavity method\cite{mpv}).  Replica symmetry breaking is associated with the existence of many states (each replica can be in one or another of these states).  A small change in the magnetic field pushes the system in a slightly 
metastable state, that may decay only with a very long time scale.  This may happens only if there are many states that 
differs one from the other by a very small amount in free energy.

  \section{Complexity} \label{SEC-Complexity}
\subsection{Metastable states}
We have seen that some important properties of glasses point to the existence of metastable states.
Although the word {\sl metastable state} has a strong intuitive appeal, we would like to  define it in a more precise way.  
There are two different (hopefully equivalent) definitions of a metastable state or valley:
\begin{itemize}
\item
From an equilibrium point of view a valley is a region of configuration space separated by the rest of the 
configuration space by free energy barriers that diverge when $N\to\infty$.  
More precisely the system, in order to go outside a valley by moving one spin (or one particle) at 
once, must cross a region where the free energy is higher than that of the valley by 
a factor that goes to infinity with $N$.
\item
From the dynamic point of view a valley is a region of configuration space where  the system remains confined, at all times smaller than an escape time  
that goes to infinity with $N$.
\end{itemize}
The rationale for assuming that the two definitions are equivalent is the 
following.  We expect that for any reasonable dynamics
\footnote{}
 where the system
evolves in a continuous way (i.e.  one spin flip at time), the system must cross a configuration of higher free energy 
when it goes from a 
valley to an other valley (this does not apply to kinetically constrained models, where some local movements are 
forbidden, so that one can find dynamical valleys that do not correspond to valleys from the equilibrium point of view). The time for escaping from a valley is given by
\begin{equation}
\tau \simeq \tau_0 \exp (\beta \Delta F)
\end{equation}
where $\Delta F$ is the free energy barrier.

It is crucial to realize that in infinite range models valleys may have a free energy density higher than that of equilibrium states.  This phenomenon is definitely not present in short range models.  Two equilibrium states with infinite mean life must have the same free energy. If there exist two phases (or valleys), denoted as $A$ and $B$, where the free energy density of $B$ is higher than that of $A$, it is clear that the system can go from $B$ to $A$ in a continuous way, by forming a bubble of radius $R$ of phase $A$ inside phase $B$.  If the surface tension between the two phases $A$ and $B$ is finite, as happens in any short range model (but not necessarily in infinite range models), for large enough $R$ the free energy of the system with a bubble of radius $R$ will be smaller than the free energy of the pure $B$ system, and the bubble will expand.  This argument does not work in mean field models where in some sense surface effects are as important as volume effects.

In the real systems there are metastable states with very 
large (e.g.  much greater than one year) mean life.  We shall consider here the {\sl 
infinite} time metastable states of the mean field approximation as precursors 
of these {\sl finite} mean life states:  corrections to 
the mean field approximation will give a finite (but large) mean life to 
these states.

\subsection{The basic definitions}
The complexity is a kind of  entropy associated with the multiplicity of metastable states.
Before discussing the difficulties related to the definition of the complexity in short 
range models, we shall present here the main definitions that are correct in the mean field approach.

The basic ideas are simple \cite{pspin,CuKu,FM,KPVI,KLEIN,FP,FRAPA,MONA}: we
partition the whole configuration  space  into valleys. If we call $Z_{\alpha}$ the contribution of each 
valley to the partition function,  the corresponding free energy is given by
\begin{equation}
Z_{\alpha}=\exp(-\beta F_\al)\ .
\end{equation}
This definition does not give us a practical way to find the valleys.  One possibility is the following.  Let us consider for simplicity of notation a monoatomic system and let us consider the density $\rho(x)$.  We can 
introduce a free energy functional $F[\rho]$ that depends on the density and on the temperature. 
The explicit form of the functional is not crucial.
We suppose that at sufficiently low temperature the functional  $F[\rho]$ has many local minima (i.e. the number of 
minima goes to infinity with the number $N$ of particles).  Exactly at zero temperature these local minima coincide with 
the local minima of the potential energy as function of the coordinates of the particles.  Let us label them by an 
index 
$\alpha$.  To each of them we can associate a free energy $F_\al$ and a free energy density $f_\al= F_\al/N$.  In this 
way the valleys are associated with local minima of the free energy functional.

In this low temperature region we suppose that the total free energy of the system can be well 
approximated by the sum of the contributions to the free energy of each particular local minimum. We thus find:
\begin{equation}
Z\equiv \exp(-\beta N f_{S}) =\sum_\al \exp(-\beta N f_\al)\ .
\end{equation}

When the number of minima is very high, it is convenient to introduce the function $\cN(f,T,N)$, i.e. the density of 
minima whose free energy density is $f_\alpha=f$.  With this notation we can write the previous formula as
\begin{equation}
Z= \int df \exp (-\beta N f) \cN(f,T,N).
\end{equation}
If we assume that $\cN$ is exponentially large in the system size, we can write
\begin{equation}
\cN(f,T,N) \approx \exp(N\Sigma(f,T)),\label{EQ_CON}
\end{equation}
where the function $\Sigma$ is called the complexity or the configurational entropy (it is the 
contribution to the entropy coming from the existence of an exponentially large number of locally 
stable configurations).

The minimum possible value of the free energy is given by $f_m(T)$, the maximum value is $f_M(T)$. The relation (\ref{EQ_CON}) is 
valid in the region $f_m(T)<f<f_M(T)$.   Outside 
this region we have that $\cN(f,T,N)=0$ for large enough $N$.  It all known  cases $\Sigma(f_m(T),T)=0$, and the function $\Sigma$ is 
continuous 
at $f_m$. On the contrary, in mean field models it  frequently happens that the function $\Sigma$ is discontinuous 
at $f_M$.

For large values of $N$ we can write
\begin{equation}
Z=\exp(-N \beta f_{S}) \approx \int_{f_m}^{f_M} df \exp (-N(\beta f- \Sigma(f,T))).\label{SUM}
\end{equation}
We can thus use the saddle point method and  approximate the 
integral  with the integrand evaluated at its maximum.
We find that
\begin{equation}
\beta f_{S}=\min_f\Phi(f) \equiv \beta f^* - \Sigma(f^*,T),
\end{equation}
where the potential $\Phi(f)$ (that will play a crucial role in this approach) is given by
\begin{equation}
\Phi(f)\equiv\beta f - \Sigma(f,T).
\end{equation}
(This formula is quite similar to the well known homologous formula for the free energy, i.e.  
$\beta 
f=\min_{E} (\beta E -S(E))$, where $S(E)$ is the entropy density as function of the energy density.)

If we call $f^*$ the value of $f$ that minimizes $\Phi(f)$, there are two possibilities:
\begin{itemize}
\item
The minimum $f^*$ is inside the interval and it can be found as a solution 
of the equation $\beta=\partial \Sigma/\partial f$.  In this case we have
\begin{equation}
\beta f_S=\Phi(f^*)=\beta f^* - \Sigma^*, \ \ \ \Sigma^*=\Sigma(f^*,T).
\end{equation}
The system at equilibrium will be found in any of the  $\exp(N \Sigma ^*)$  minima which have the free energy density $f^*$.
The total entropy of the system is thus the sum of the entropy of a typical minimum and of 
$\Sigma^*$, i.e. the contribution to the entropy coming from the exponentially large 
number of microscopical configurations.

\item
The function $\phi(f)$ reaches its minimum $f^*$ at the extreme value of the range of variability of $f$.  Then $f^*=f_m$ and $\Phi=f_m$.  In 
this 
case the contribution of the complexity to the total free energy is zero.  The relevant states 
all have the free energy density $f_s$, their number does not grow exponentially with $N$. They have a difference in free energy density that is of order $N^{-1}$ (a difference in total free energy of order 
1).  From the point of view of replica theory, this phase where the free energy is dominated by a few different minima is called the replica 
symmetry broken phase \cite{mpv,parisibook2}.
\end{itemize}

This discussion shows that all the properties  of the system depend crucially on the free 
energy landscape \cite{ACP}, i.e. the function $\Sigma(f,T)$, the distance among the 
minima, the height of the barriers among them...

\subsection{Computing the complexity}\label{CC}

The complexity is the entropy of metastable states. We need a method that allows to estimate it, without having to solve the impossible task of finding all the (many!) minima of the free energy functional.  The solution consists in using the system as a polarizing tool for itself, by introducing new artificial couplings. This new approach also works in cases where the free energy functional is not known 
exactly, so that its minima cannot be computed.

The basic idea  is to start from an equilibrium configuration and to explore the configuration 
space phase around it \cite{FP,FRAPA,MONA,Me,MePa1,MeAspen}. 

More precisely, we study a system of $N$ interacting atoms defined by their positions $x=\{x_i\}$, $i=1,...,N$, with a Hamiltonian $H(x)$, 
which might be for instance a pairwise interaction potential
\begin{equation}
H(x)= \sum_{i<j} V_{ij}(|x_i-x_j|)\ ,
\end{equation}
like either a hard sphere potential,
a `soft sphere' potential ($V_{ij}(r)=A/r^{12}$), or a Lennard-Jones
 potential ($V_{ij}(r)=A/r^{12}-B/r^6$).
Given two configurations $x$ and $y$ we define their overlap as
$
q(x,y) = -1/N \sum_{i,k=1,N} w(x_{i}-y_{k}),
$
where $w(r)$ is an arbitrary function with the following properties. $w$ increases with the distance, $w(r)=-1$ for $r$ small, $w(r)=0$ for $r$ larger than the typical interatomic distance.

We now add an extra term to the Hamiltonian:
we define
\begin{equation}
\exp (-N \beta F(y,\eps))=\int dx \exp( -\b H(x)+ \beta  \eps N q(x,y))\  .
\end{equation}
Now suppose that $y$ is an equilibrium configuration at the inverse temperature $\beta$, and define
\begin{equation}
\Gamma(\eps)= \lan F(y,\eps)\ran_{\sig}\ ,
\end{equation}
where $\lan \cdot \ran_{\sig}$ denotes the average over $y$, with the measure $(1/Z) \exp(-\b H(y))$. For $\eps$ sufficiently large, the 
new term in the Hamiltonian forces the system $x$ to be near to the reference system $y$, 
and produces a \emph{quenched} disorder for the system $x$.  By changing the value of $\eps$ we can explore 
the 
phase space at a given distance around a given equilibrium configuration $y$.  In the end we average over $y$ the logarithm of the 
$y$ dependent free energy.
We can measure the typical overlap between the configuration $x$ and its reference configuration $y$ by
\begin{equation}
q(\eps)=-\lan \frac{\partial F(y,\eps)}{\partial \eps}\ran_{\sig}\ .
\end{equation}

The quantity $\Gamma(\eps)$ is well defined and it may be computed  also in numerical simulations.  However it is 
interesting to evaluate it in mean field models, where analytic computations are possible.  The analytic computation of 
$\Gamma(\eps)$ can be done by considering $n=1+s$ replicas. We introduce $n$ configurations $x^a$, $a=1,\dots,n$, each of them being a configuration of $N$ particles: $x^a=\{x_i^a\}$ , $i=1,...,N$. These replicated systems interact through the Hamiltonian:
\begin{equation}
H_{s}(x^1,\dots,x^n)\equiv H(x^1)+\sum_{a=2}^{1+s} H(x^a)-\eps N \sum_{a=2}^{1+s}q(x^1,x^a)\ .
\end{equation}
Here $x^1$ plays the role of $y$ and the  $x^a$ (for $a=2,1+s$)  are 
$s$ replicas of the $x$ variables.
The quenched limit (where there is no feedback reaction of the $x$ variables on 
the $y$ variables) is obtained in the limit $s \to 0$. It gives:
\begin{equation}
\Gamma(\eps)= -\frac{1}{N\beta} \lim_{s \to  0} \frac{\partial }{ \partial s} \ln \ato 
\int \prod_{a=1,1+s} dx^a \exp(- \b H_{s}(x^1,\dots,x^{1+s})) \cto
\end{equation}
 It is 
convenient to define the Legendre transform of $\Gamma(\eps)$, defined as
\bea
W(q)=\Gamma(\eps(q))+\eps(q) q \ , \\
\frac{\partial W(q) }{ \partial q} =\eps(q) \ .
\eea
The potential $W(q)$ has the meaning of the free energy with  the constraint that the 
overlap of our configuration $x$ with the generic thermalized configuration $y$ is equal to $q$:
defining
\begin{equation}
\exp (-N \beta W(y,q))=\int dx \exp( -\b H(x)) \delta(q- q(x,y))\  ,
\end{equation}
 we have $W(q)= \lan W(y,q)\ran_{\sig}$.
As far as we are interested in studying the $q$-dependence of $W(q)$, we can shift the origin of $W$ so that $W(0)=0$.

\begin{figure}
 \includegraphics[width=.6\textwidth]{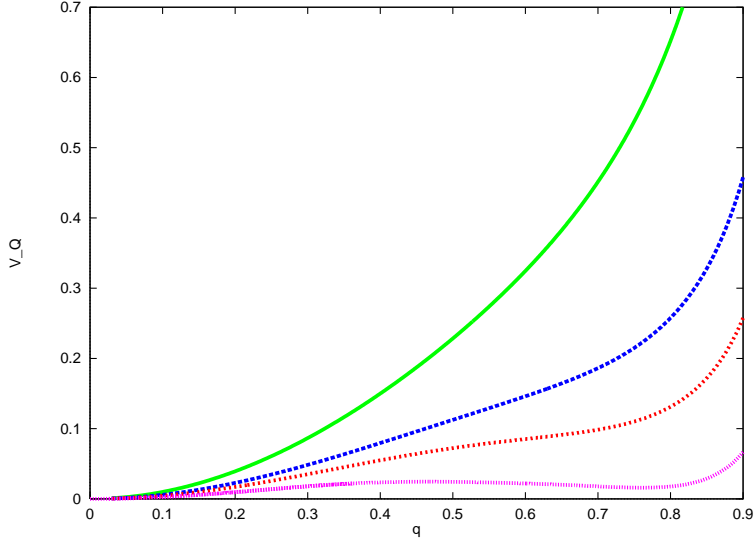}
\caption[0]{\protect\label{W}
Different shapes of the function $W$ for various temperatures: the upper curves correspond to higher temperatures.
  }
\end{figure}

In the following we will study the phase diagram of the model in
the $\eps -T$ plane. Exact computations can be found in the literature mainly in the mean field models \cite{FP,FRAPA}, but the conclusions have a general validity.
If one computes the functions $\Gamma(\eps)$ and $W(q)$ in a mean field model, one typically finds that the shape of 
the 
function $W$ is characteristic of a mean-field system undergoing a first order phase transition.  At high enough 
temperature $W$ is an increasing and convex function of $q$ with a single minimum for $q=0$.  Decreasing the 
temperature 
below a value $T_f$, where for the first time a point $q_f$ with $W''(q_f)=0$ appears, the potential looses the 
convexity property and a phase transition can be induced by a field.  A secondary minimum develops at $T_d$, the 
temperature of the `dynamical transition' \cite{KTW}, signaling the presence of long-life metastable states.  The height of 
the secondary minimum reaches the one of the primary minimum at $T=T_s$ and thermodynamic coexistence at $\eps=0$ takes 
place.  This is the thermodynamic transition.  In figure \ref{W} 
we show the shape of the potential in the various regions.

\begin{figure}
 \includegraphics[width=.6\textwidth]{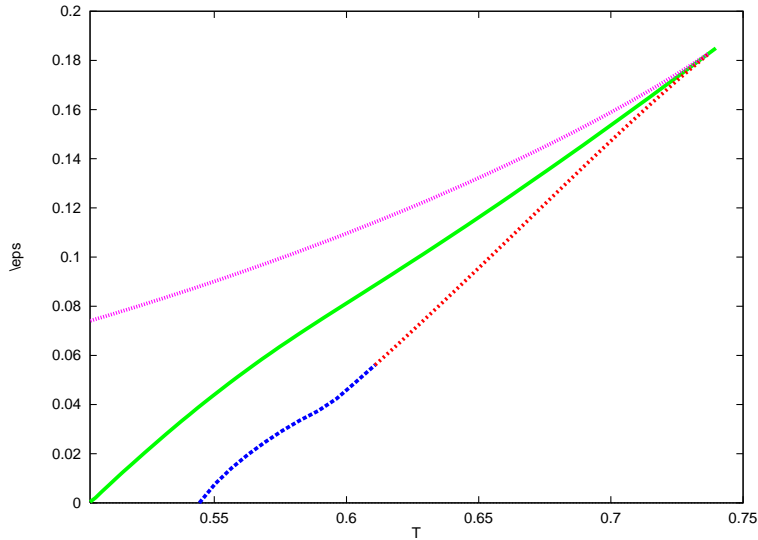}
  \caption{\protect\label{F_E1}
Phase diagram in the $T-\eps$ plane.   At the upper curve the low $q$ solution disappear, at the lower curve the high 
$q$ 
solution disappear and two locally stable solutions are present only in the region among the upper and lower curves.  
The middle curve is the coexistence line where the two solutions have equal free energy.  The coexistence line touches the 
axes $\eps=0$ at $T=T_s$, while the lower curve touches it at $T=T_{D}$.  }
\end{figure}

Therefore the potential $W(q)$ has usually a minimum at $q=0$, where $W(0)=0$.  It may have a secondary minimum at 
$q=q_D$.  Depending on the value of the temperature, we meet a few different situations:
\begin{itemize}
    
\item At $T>T_D$ the potential $W(q)$ has only the minimum at $q=0$. The dynamical transition temperature is defined as
$T_D$.  A more careful analysis \cite{BAFRPA} shows that for $T_D<T<T_V$ there are still valleys with energy {\sl less} 
than the equilibrium one, but these valleys cover a so small region of phase space that they are not relevant for 
equilibrium physics.

\item Exactly at $T=T_D$ we sit at a phase transition point where some susceptibilities are divergent.  This fact 
implies (in short range models) that there is a divergent dynamical correlation length that is related to dynamical 
heterogeneities \cite{corr_length}.

\item At $T_c<T<T_D$, there is a secondary minimum of $W(q)$ with a positive value, $W(q_D)>0$.  In this intermediate temperature region,
 we can put 
one replica $y$ at equilibrium and have the second replica $x$ in a valley near to $y$.  It happens that the internal 
energy of both the $y$ configuration (by construction) and of the $x$ configuration are equal to the equilibrium 
one.  However the number of valleys is exponentially large (there is a finite complexity) so that the free energy of a single valley is higher than 
the total free energy.  One finds in this way that $W(q_D)>0$ is given by
\begin{equation}
W(q_D)= \frac{\ln \N _e }{ N} \equiv T \Sigma^{*}
\end{equation}
where $ \N _e $ is the average number of the valleys having the equilibrium energy \cite{MONA,PP} . The total entropy is 
\be \label{EQ_CON}
S=\Sigma^*+S_v\, ,
\ee
 where $S_v$ is the internal entropy of one valley. The complexity $\Sigma^*$ is thus the difference between the full entropy and the internal valley entropy.
It vanishes at $T_c$ and becomes exactly equal to zero for $T<T_c$ \cite{KTW} .

\item In the whole region  $T<T_c$, the secondary minimum at $q_D$ is at the same level as the minimum at $q=0$: $W(q_D)=0$.  This means that 
 we can put two replicas both at 
overlap $0$ and at overlap $q_{EA}$ without paying any prize in free energy.  In this case $q_D$ is the Edwards-Anderson order parameter,
$q_D=q_{EA}$.
\end{itemize}

Although the behavior of this potential function $W$ is analogous to the one found in ordinary systems undergoing a first 
order phase transition, the interpretation is here radically different (a hint of the existence of a basic difference is the fact that $W(q_D)=0$ in the whole phase $T<T_c$, and not just at $T=T_c$).  While in ordinary cases different minima 
represent qualitatively different thermodynamical states (e.g. gas and liquid), this is not the case here.  In our 
problem the secondary  local minimum of $W(q)$ found at $q=q_D$  appears when ergodicity is broken, and the configuration space splits into an exponentially 
large number of components.  The two minima are different manifestations of states with the same characteristics.  The 
height $W(q_D)$ of the secondary minimum, relative to the one at $q=0$ measures the free-energy loss to keep the system near one 
minimum of the free energy (in configurations space).  This is just the complexity $T\Sigma$, i.e. the logarithm of the 
number of distinct valleys of the system.
\begin{figure} \label{FIG_W}
    \includegraphics[width=.6\textwidth]{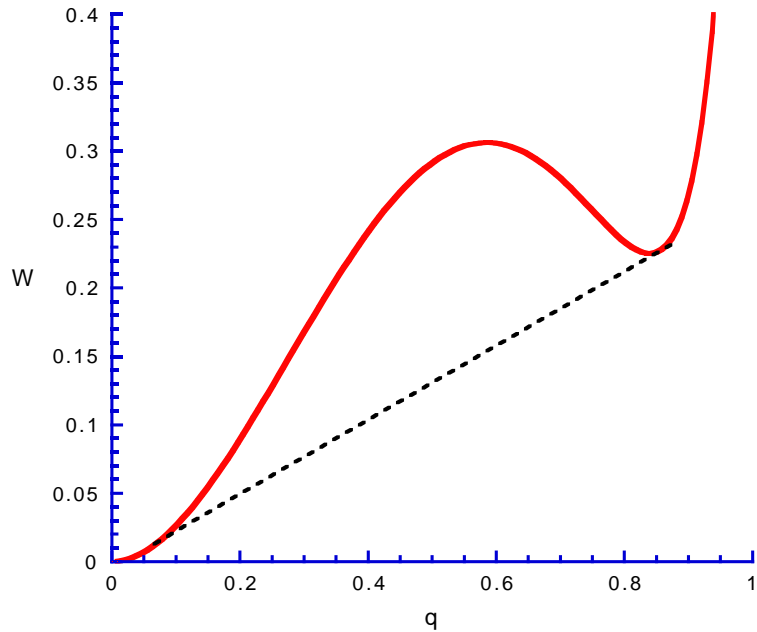}
 \caption[0]{\protect\label{MAX}
The full line is the function $W(q)$  computed in the mean field approximation. The dashed line is the correct
result (Maxwell construction).}
\end{figure}

It is interesting to study the overlap function $q(\eps)$. This function gives the typical overlap of
the configuration $x$ with its reference configuration $y$, for a given value of the coupling $\eps$. Its value is given by the solution
of
the equation $\partial W(q)/ \partial q =\eps$.  In the region of the $T-\eps$ plane shown in fig. \ref{F_E1}, this equation  has two stable
(and one unstable) solutions (the stable solutions are the ones which correspond to a local minimum of 
$W(q)-\eps q$). Along the upper and lower curves of this figure, one of 
the 
two solutions loose its stability and  disappears: these two curves are the equivalent of the spinoidal lines in 
usual first order transition. The point where the lower curve crosses the axis $\eps=0$ is the dynamical transition 
\cite{FRAPA}: 
only at lower temperatures can the two systems remain with a high value of the overlap 
without having a force that keeps them together (i.e. $\eps=0$). On the contrary the static transition is 
characterized by the fact that the coexistence line touches  the axis $\eps=0$.

\begin{figure}
     \includegraphics[width=.6\textwidth]{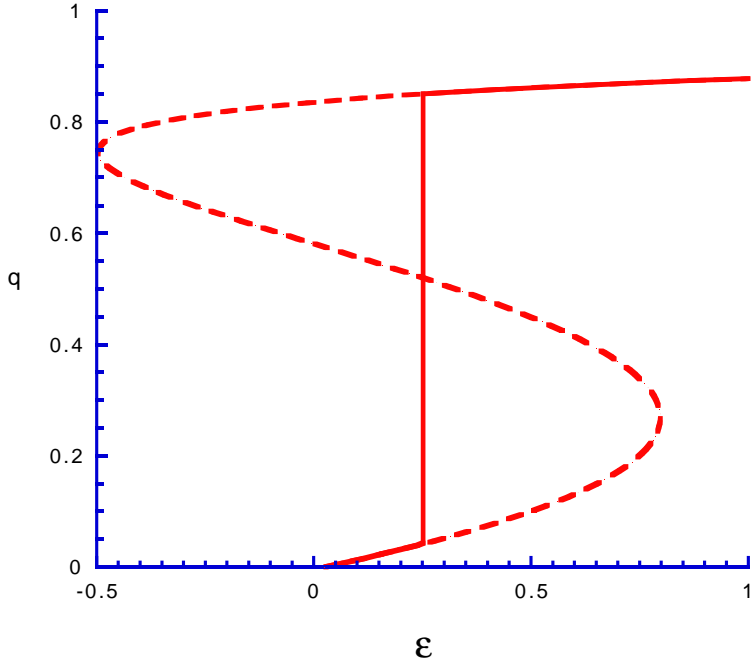}
 \caption[0]{\protect\label{qeps}
The shapes of the function $q(\eps)$ for $T_c<T<T_{D}$ in a finite range system: the full line is the correct result and the dashed line
is the output of a mean field approximation.
  }
\end{figure}

In systems with finite range interactions, the situation must be considered more carefully.
General arguments tell us that the free energy is a convex function of the $q$, so that the correct shape of the 
function $W$  can be obtained by the Maxwell construction (see fig.  \ref{MAX}). This means that, when 
$T_c<T<T_{D}$, the equilibrium overlap function $q(\eps)$ has the monotonously increasing  behaviour  shown in Fig. \ref{qeps}. 
We see a typical metastability effect: when one decreases $\eps$, the overlap starts from a large value close to $q_D$. At a critical value of 
$\eps$, given by the vertical line in Fig. \ref{qeps}, the equilibrium overlap jumps to a value close to $0$. However there exists a metastable branch
at large overlaps, and it is reasonable to believe that, if the rate of decrease of $\eps$ is not infinitely small, the system will follow this metastable branch. 
The mean-field computation of the complexity, giving $\Sigma^*=W(q_D)/T$, corresponds to the minimum of $W(q)-\eps q$ at $\eps=0$ and $q=q_D$. 
As can be seen from the figures,  for $T_c<T<T_{D}$, where $\Sigma^*$ is non-zero, this point is always in 
the metastable region.  This causes an intrinsic ambiguity in 
the definition of complexity in finite range systems: the free energy is a notion defined in a metastable phase and as such it 
is not defined with infinite precision.  
However we can use the fact that the free energy is a $C^{\infty}$ function of $\eps$ near the discontinuity point to 
extrapolate from high $\eps$ to $\eps=0$. The extrapolation allows to obtain the free energy in the metastable region, and to compute 
$\Sigma^*$.  The  ambiguity created by this extrapolation becomes smaller when $T$ becomes closer to $T_c$ (the amount of the extrapolation becomes smaller) and in general it is 
rather small 
unless we are very near to the dynamic phase transition.  So this ambiguity is not important from practical 
purposes; however it implies that there is no sharp, infinitely precise definition of the 
equilibrium complexity.  If we forget this intrinsic ambiguity in the definition of the complexity we may arrive to 
contradictory results.

\subsection{Complexity and replicas}\label{MORE}
The complexity function plays a major role in this whole approach. Let us see how it can be computed.
As we have seen, when a system can be in many valleys, we can write
\begin{equation}
Z(\beta)=
\sum_{a} \exp( -\beta N f_{a}(\beta))= 
\int d \cN(f,\beta) \exp (-\beta N f) \ ,
\end{equation}
where $f_{a}(\beta)$ is the free energy density of the valley labeled by $a$ at the temperature $\beta^{-1}$, and 
$\cN(f,\beta)$ is the number of valleys with free energy density less than $f$, which is supposed to be, in some regions of the
$f,\beta$ space, exponentially large: $\cN(f,\beta) =\exp(N \Sigma(f,\beta))$. The complexity 
$\Sigma(f,\beta)$ is supposed to be positive in the region $f>f_{m}(\beta)$ and to vanish continuously at $f=f_{m}(\beta)$.  The quantity $ 
f_{m}(\beta)$ is the minimum value of the free energy: $\cN(f,\beta)$ is zero for $f< f_{m}(\beta)$.

A simple strategy to 
compute 
the complexity $\Sigma(f,\beta)$ is the following.  We introduce a modified partition function in which the
various states are weighted with an inverse temperature $\gamma$ which may differ from $\beta$. The 
modified partition function is defined as 
\begin{equation}
Z(\gamma;\beta)\equiv \exp ( -N \gamma G(\gamma;\beta))=\sum_{a} \exp( -\gamma N 
f_{a}(\beta)).
\end{equation}
It is evident that $Z(\beta;\beta)$ is the usual partition function and $G(\beta;\beta)$ is the usual free energy.
Using standard saddle point arguments it can be easily proven that in the limit $N 
\to \infty$, for a given value of $\beta$, the function $G$, considered as a function of $\gamma$, is the Legendre transform of 
$\Sigma$, seen as a function of $f$. Precisely :
\be
\gamma G(\gamma;\beta)=  \gamma f - \Sigma(\beta,f),\ \ \ 
f=\frac{\partial (\gamma G(\gamma;\beta) )}{ \partial \gamma}.
\label{LTbg}
\ee
The complexity is obtained from $G(\gamma;\beta)$ in the same way as the 
entropy is obtained from the usual free energy \cite{MONA,LJ,Me,MePa1}:
\begin{equation}
\Sigma(\beta,f)=\gamma^2 \frac{\partial G(\gamma;\beta)}{\partial \gamma}\ .
\end{equation}

A few observations are in order:
\begin{itemize}
\item In the new formalism, the parameter $\gamma$, the free energy, the complexity, and the function $G$, play respectively 
the same role as $\beta$, the energy, the entropy, and the free energy in the usual formalism.  

\item In the new formalism the usual inverse temperature $\beta$ only indicates the value of the temperature that is 
used to compute the free energy and the new inverse temperature $\gamma$ controls which part of the free energy 
landscape  is sampled.  

\item When $\beta \to \infty$ (at least in mean field models) the new formalism samples the energy landscape: 
\begin{equation}
Z(\gamma;\infty)=\sum_{a} \exp( -\gamma N e_{a})=\int \nu(e) de \exp (- \gamma N e)
\end{equation}
where $e_{a}$ are the minima of 
the Hamiltonian and $\nu (e)$ is the density of the minima of the Hamiltonian.

\item The equilibrium complexity is obtained by taking $\gamma=\beta$. On the other hand its value in the limit $\gamma=\infty$ 
give us information on the number of minima of the Hamiltonian.
\end{itemize}

In principle it is possible to get the function $\Sigma(f)$ by computing directly the number of solution of the TAP 
equations for a given value of the free energy density.  However it is simpler to obtain it by using the replica 
formalism and it is reassuring that one gets the same results with both methods \cite{MONA,FRAPA,PP,Me,CGG,CGMP}.

\begin{figure}
\includegraphics[width=.5\textwidth,angle=-0]{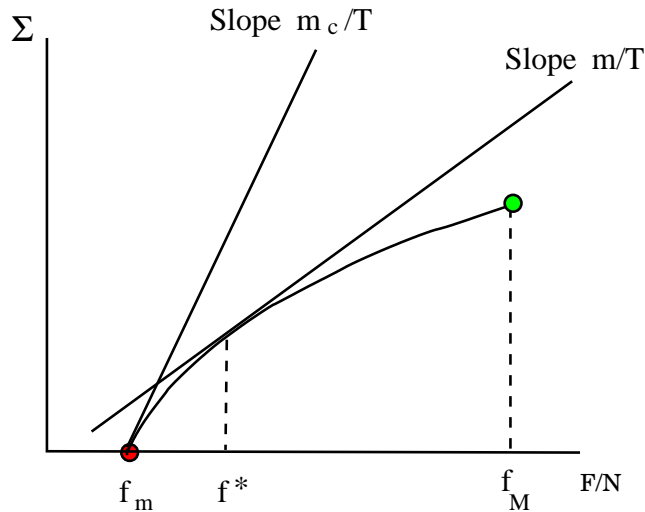}
\caption{The  curve shows a typical shape of the complexity $\Sigma(\beta,f)$ as function of $f$. The thermodynamic properties of the system with $m$ replicas is dominated by  the states free energy densities $f^*(m)$ such that $\frac{\partial \Sigma}{\partial f}(f^*)=\frac{m}{T}$. When $m$ increases beyond a critical value, solution of $\frac{\partial \Sigma}{\partial f}(f_m)=\frac{m}{T}$, the dominating states are those with the lowest possible free energy density, $f=f_m.$ }\label{smqualit}
\end{figure}

The computation of the modified partition function $Z(\gamma;\beta)$ can be easily done in 
the replica formalism \cite{FRAPA,PP,Me}.  If we consider a system with $m$ replicas (with $m$ integer) and 
we constrain them to stay in the same state we find that
\begin{equation}
Z(\beta,m)=
\sum_{a} \exp( -\beta m N f_{a}(\beta))
\end{equation}
This expression coincide with $Z(\gamma;\beta)$, where the new inverse temperature is
\begin{equation}
\gamma=m \beta \ .
\end{equation} 
Therefore there is 
a very simple method to compute  $G(\gamma;\beta)$. We must consider the partition 
function of $m$ replicas that are constrained to stay in the same state. This means that the overlap between the various replicas 
must be a value $q$ which is large enough (as we will see, $q$ must be chosen in a self consistent way).
We shall study this replicated system at a given temperature, varying the value of $m$. The partition function is
written in terms of the complexity as:
\begin{equation}
Z(\beta,m)=e^{-N\beta G(m,\beta)}= \int df e^{N(\Sigma(\beta,f)-\beta m f)}\ .
\end{equation}
It is thus dominated by free energy densities $f^*(m)$ such that $\frac{\partial \Sigma}{\partial f}(f^*)=\frac{m}{T}$, and $G$ as function of $m$
is the Legendre transform of $\Sigma$ as function of $f$. By varying $m$ at a fixed temperature, we can thus reconstruct the 
complexity function. Fig.\ref{smqualit} shows the typical behaviour of this Legendre transformation, when the temperature is $T<T_d$ (this curve has been obtained using a spin glass model with $p$-spin interactions, but as we shall see in the following sections the qualitative behavior is the same in all glass models, at the mean field level). When $m$ is small enough, the dominating free energy  $f^*$ is in a region where $\Sigma(f^*,T)$ is positive. When $m$ increases, there is a phase transition when the typical free energy density $f^*(m)$ reaches the minimum free energy $f_m$ (when  $\frac{\partial \Sigma}{\partial f}(f_m)=\frac{m}{T}$). For larger values of $m$, the typical configurations remain those at $f=f_m$, and the total free energy density does not vary with $m$.

Although we have based our discussion on mean-field models, we expect that the qualitative features of the phase 
diagram
presented here survive in finite dimension. The existence of a coexistence line, terminating in a critical 
point, is a constitutive feature of systems whose physics is dominated by the existence of long lived metastable states 
like glasses.  As we shall see below, the predictions derived from the replica approach 
can be submitted to numerical tests in glassy model systems like e.g. 
Lennard-Jones or hard spheres, or polymer glasses.  For example the identification of the complexity $\Sigma$ as the 
free energy difference between the stable and the metastable phases provides another method to measure this quantity in 
a 
simulation.  Indeed the ending of the transition lines in a critical point implies that the metastable state can be 
reached via closed paths in phase diagram leaving always the system in (stable or metastable) equilibrium; the free 
energy difference of the two phases can be computed integrating the derivative of the free energy along such a closed 
path.

\subsection{Old replicas and new replicas}
It is interesting to have an understanding of the relationship between the new use of replicas (sometimes named `clones' in this context \cite{Me})
 that we are proposing here
and the `old'  notion of replicas used in the study of disordered systems. We have already stressed the very different philosophy, as the use of replicas that we have presented in the above sections does not assume the existence of any quenched disorder. This very different general philosophy explains why the replica approach, so successful in the study of spin glasses\cite{mpv}, has appeared in the science of structural glasses only many years later \cite{MPR,BM94,FH,CKPR,CKMP}. Yet,
in practice, the two uses of replicas are very deeply related. This explains why the vocabulary and concepts of replica symmetry breaking have been immediately transcribed to the study of structural glasses.

 In order to understand this important point, let us consider a system with quenched disorder, described by a Hamiltonian $H(J,s)$, where $s$ are the thermalized variables (they could be spins in a spin glass, or position variables in a structural glass model with quenched disorder), and the $J$ are quenched random variables, distributed according to a law $P(J)$. 
A given sample means a given set of $J$ variables. For each given sample the $s$ variables are thermalized, meaning that their distribution
is
\begin{equation}
P_J(s)=\frac{1}{Z_J}e^{-\beta H(J,s)}\ .
\end{equation}
The  thermodynamic properties are obtained from the partition function $Z_J$ and the free energy  density $F_J=-\log Z_J/(N\beta)$.
The self-averaging property, typical of systems with quenched disorder, means that, while $Z_J$ has large sample-to-sample fluctuations,
$F_J$ is self-averaging, meaning that the distribution of $F_J$ concentrates around its mean value in the large $N$ limit. This indicates that, for large samples, almost all samples will have a free energy density equal to $F=E_J F_J= \sum_J P(J) F_J$. Therefore the thermodynamic properties are obtained from an estimate of the quenched average:
\begin{equation}
e^{-\beta N F}=E_J \log Z_J = \sum_J P(J) \log\left( \sum_s e^{-\beta H(J,s)}\right)\ .
\end{equation}
The usual replica method is based on the observation that, in general, it is very difficult to perform this quenched average,  but it is often relatively easy (at least in mean field problems) to compute the average of a system which has been replicated $n$ times, with $n$ a positive integer.
One thus introduces the replicated partition function
\begin{equation}
 Z_J^n= \sum_{s_1,\dots,s_n}e^{-\beta \sum_{a=1}^n H(J,s_a)}\ .
\end{equation}
Taking the average over samples gives 
\begin{equation}
 E_J Z_J^n=\sum_J P(J)  \sum_{s_1,\dots,s_n}e^{-\beta \sum_{a=1}^n H(J,s_a)}= \sum_{s_1,\dots,s_n}e^{-\beta H_n(s_1,\dots,s_n)}\ ,
\end{equation}
which has become a problem of $n$ interacting replicas (the interaction $H_n$  is induced by the average over $J$) without any quenched disorder.
Then the usual replica method proceeds by using 
\begin{equation}
  E_J\log Z_J  =\lim_{n\to 0} \frac{E_J Z_J^n-1}{n}\ .
\end{equation}
If one knows how to estimate $E_J Z_J^n $  at $n$ close to $0$, this gives the desired result.

In many mean field models, the  average $E_J Z_J^n $ is obtained by a saddle point procedure in terms of an overlap matrix $Q_{ab}$,  $1\le a \le b\le n$, where $Q_{ab}$ is a suitably defined overlap between the two replicas $a$ and $b$. A much studied class of problems are those displaying `one-step replica symmetry breaking', where the equilibrium glass phase (obtained at $T<T_c$) is obtained for an overlap matrix
\begin{equation}
 Q_{ab}=\left\{\begin{array}{l} q_1 \ \ \ if \ \ \ I[a/x]=I[b/x] \\
     q_0 \ \ \ if \ \ \ I[a/x]\neq I[b/x]\end{array}\right. 
\end{equation}
where $I[p]$ is the integer part of $p$. This means that the set of $n$ replicas is partitioned into $n/x$ groups, each containing $x$ replicas. The overlap between two replicas in the same group is $q_1$, the overlap between two replicas in different groups is $q_0<q_1$. 
In mean field models, the values of $q_0, q_1$ and $x$ are obtained as the solutions of the stationarity condition of the free energy expressed in terms of $q_0,q_1,x$ (and, when $n \to 0$, the parameter $x$ must be in the interval $0<x<1$)\cite{mpv}.

This structure is the simplest example of the hierarchical, `ultrametric', structure which is typical of replica symmetry breaking  \cite{MPSTV}. It appears in particular in the random energy model \cite{NICOLA, REM,GROMEZ}, and in many other spin glass models with $p$-spin interactions \cite{pspin}. These are precisely the systems
which have a phenomenology close to the one of structural glass formers \cite{KTW} (the corresponding one step replica symmetry breaking transition 
is also called a random first order transition). 

Let us understand how our new replica approach, described in the previous section, applies to these systems exhibiting a one step replica symmetry broken phase. Take a given sample $J$. We should introduced $m$ coupled copies of the systems, call them $s_a$, with $a=1,\dots, m$. These copies should be constrained  in such a way that their overlap takes the large value, which is here equal to $q_1$. Therefore the new-replicated partition function and free energy, for this given sample, is expressed as 
\begin{equation}
 e^{-\beta N G(m,\beta,J)}=Z_J(m)=  \sum_{s_1,\dots,s_m}e^{-\beta \sum_{a=1}^m H(J,s_a)}\prod_{a<b}\delta(Q_{ab},q_1)\ .
\end{equation}
From self averageness, one expects that the distribution of $G(m,\beta,J)$ concentrates around its mean, so that we can compute its typical value through the mean $E_J G(m,\beta,J)$. This we can do through the introduction of $n'$ `old-type' replicas, in the limit where $n'$ goes to zero:
\begin{equation}
  G(m,\beta)=E_J G(m,\beta,J) =-\frac{1}{\beta N}\lim_{n'\to 0} \frac{E_J Z_J(m)^{n'}-1}{n'}\ .
\end{equation}
At this stage, we see that if we write $n'=n/m$, the new formalism reduces exactly to the old one: the $n' m=n$ replicas are grouped into $n'=n/m$ groups of $m$ replicas, where the intra-group replica overlap is constrained to take the value $q_1$. The saddle-point analysis of the model will then naturally lead to the fact that the inter-group overlap takes the value $q_0$, as in the old approach. The parameters $x$ and $m$ play exactly the same role in the two approaches, and the quenched free energy $F(m)$ for a given value $x=m$ is related to the quenched free energy $G(m,\beta)$ by $F(m)=G(m,\beta)/m$.

The reader should now be puzzled by the fact that the new approach can be done for any value of $m$, while in the old approach one chooses the value of $x$ where the free-energy is maximal.  In the old approach one obtains the quenched free energy density $F$ by choosing the value of $x$ which makes it stationary: $\partial F /\partial x=0$. As $F(x)=G(x,\beta)/x$, this condition amounts to saying that $G(x,\beta)=x \partial G(x,\beta)/\partial x $, and the Legendre transform formulas (\ref{LTbg}) show that this amounts to fix the value of $x=m$ exactly such that the free energies of the valleys $f$ is the minimal one, $f=f_m$, and therefore the complexity $\Sigma(f_m)$ is zero. This is actually the correct condition: in the equilibrium glass phase, for $T<T_c$, we have seen that the dominating valleys are precisely those with $f=f_m$. So the new replica approach is able to derive all the results of the old replica approach, with the extra bonus that the interpretation of the condition  $\partial F /\partial x=0$, which is always somewhat mysterious in the old approach, now acquires a very clear meaning: this is the condition insuring that one selects the valleys with free energy density $f=f_m$.

There is one point that should be stressed. In systems with quenched disorder, like spin glasses, the free energy per replica depends on the total number of replicas. The average over disorder induces an interaction among replicas. In systems like structural glasses, where quenched disorder is absent, the free energy per replica does not depend on the total number of replicas and replicas do not interact. However also in this case, when we constrain the replicas to be near one to the other, replicas do interact and under this condition we have a dependance of the the free energy per replica on the total number of replicas. 

This phenomenon, present in the glassy phase, is similar to spontaneous symmetry breaking: there is no interaction among replicas in the Hamiltonian, but adding and infinitesimal coupling among replicas we end up with a system with strong correlations among replicas. This is a quite general phenomenon that happens each time there is a spontaneous symmetry breaking of the usual kind. For example if we take two replicas of a ferromagnetic model at low temperature and we couple them together (albeit infinitesimally) we end up with the magnetization pointing in the same direction in the two replicas. While we don't need this construction in a ferromagnetic case, where we know that the ordered phase has a spontaneous  magnetization,  it becomes very very useful in disordered systems where we do not know the structure of the ordered phase (this was the original motivation\cite{EA} for introducing the Edwards-Anderson order parameter $q$).

\subsection{A summary of the results}
Let us now summarize the results. As we have seen we can distinguish a few temperature regions.
\begin{itemize}
	
\item For $T>T_f$ the only minimum of the free energy functional is given by the high temperature result: we call it the 
liquid minimum (in the spins models it corresponds  to a zero magnetization phase).

\item For $T_f>T>T_D$ there is an exponentially large number of minima \cite{KPVI,BAFRPA,PP}.  For some values of the 
free energy density the complexity $\Sigma$ is different from zero, however the total contribution to the free energy coming 
from these minima is higher that the one coming from the liquid solution with zero magnetization.  The value $T_{D}$ coincides with the critical temperature of the mode coupling approach and in 
the glass community it is often called $T_{c}$.   The real critical temperature of the model, that we call $T_{c}$, is usually 
called $T_{K}$ in the glass literature.

\item The most interesting situation happens in the region where $T_D>T>T_c$ (or 
$T_c>T>T_K$ using the glassy notation).  In this region the free energy is still given by the 
high temperature solution (with zero magnetization in spin models).  It is extremely surprising \cite{FP,MONA} that the 
free 
energy in this phase can also be written as the sum of the contributions of an exponentially large number of non-trivial 
minima as in eq. \ref{SUM}).

Although the free energy is analytic at $T_D$, below this temperature  the system at each given moment may stay in 
one of the exponentially large number of minima.  The time ($\tau$) to jump from one minimum to an other minimum is 
quite large and it is controlled by the height of the barriers that separate the different minima.  In the mean field 
approximation (i.e. for infinite range models) this height diverges with the system size, it is proportional to $\exp (A N)$ with some 
non-zero $A$.  In short range 
models in finite dimensions we expect that the barriers are finite and $\tau\approx \tau_{0} \exp (\beta \Delta(T))$.
The quantity $\beta \Delta(T)$ is often a large number also at the dynamical temperature \cite{CGGP}  (e.g. O(10))
and the correlation time will become very large below 
$T_D$ and for this region $T_D$ is called the dynamical transition point.  The correlation time (that should be 
proportional to the viscosity) should diverge at $T_{K}$.  The precise form of this divergence is not completely
understood.  It is natural to suppose that we should get a divergence of the form $\exp(A/(T-T_{K})^{\nu})$ for an 
appropriate value of $\nu$ \cite{VF}. Several attempts at studying this problem 
can be found in \cite{KTW,KLEIN,BiBo04,franz_nuc}.  

The equilibrium complexity is different from zero (and is of order 1) when the 
temperature is equal to $T_D$;  it decreases when the temperature decreases and it vanishes 
linearly at $T=T_c$.  At this temperature (the so-called Kauzmann temperature) the entropy of a 
single minimum becomes equal to the total entropy and the contribution of the complexity to the total 
entropy vanishes. At an intermediate temperature $T_{g}$ the correlation time becomes so large that it cannot be 
observed any more by humans.
\item
In the region where $T<T_c$ the free energy is dominated by the contribution of a few minima of the 
free energy having the lowest possible value.  Here the free energy is no longer the analytic 
continuation of the free energy in the fluid phase.  A phase transition is present at $T_c$ and the 
specific heat is discontinuous here. 
\end{itemize}

\subsection{Comments on some criticisms}

The existence of a finite complexity density in finite dimensional systems  with short range forces has been criticized by many authors. In the nutshell they criticism arise from the well know observation that equilibrium metastable state cannot exist in finite-dimensional systems  with short range forces. This is obvious: in this situation the barrier is finite and therefore the supposed metastable state would decay; moreover it is not possible to define in a natural way (e.g. respecting the local equilibrium Dobrushin-Lanford-Ruelle relations) a Boltzmann weight that is concentrated on the metastable state.

If we had a finite equilibrium complexity, the free energy density of each of the exponential large number of equilibrium states would be equal to the free energy  plus the complexity. Such a putative state would have a larger free energy  than the equilibrium one and therefore it cannot be a real thermodynamic equilibrium state. Of course this argument, proving the non-existence of a finite complexity density, can be formulated in many different ways.

The tentative of defining a complexity by counting the number of inherent structures (i.e. minima of the Hamiltonian) does not work. Indeed there are two possibilities:
\begin{itemize}
\item One considers all the minima of the Hamiltonian. However this correspond to counting as different two minima that differ by the position of a single atom, i.e. a single localized defect. The fact that the number of minima of this kind is exponentially large with the volume is a quite common phenomenon which is also present in crystalline systems as soon as localized defects are present. However it would be witless to consider configurations that differ by a defect as different equilibrium state and in any case the barrier for going from one state to an other state would be finite.
\item One can try to regroup in the same state different  minima of the Hamiltonian that differ one from the other by a localized change. This is more sensible, however a careful analysis shows that the resulting complexity cannot be proportional to the volume, for reasons very similar to those that lead to the non-existence of a finite thermodynamic complexity density.
\end{itemize}

The previous criticism are absolutely correct, however they apply to a {\em sharp} definition of complexity
but not to the more sophisticated 
definition presented above, where the complexity density is defined in a {\em fuzzy} way. 

Indeed we have seen in the discussion 
at the end of Sect.(\ref{FIG_W}) that the determination of the complexity contains an intrinsic ambiguity, 

as it is obtained by the continuation of a free energy into a metastable region. As always happens we can define an approximate value of the free energy in a metastable state as function of the observation time $\tau$. Although the free energy must be the equilibrium one in the infinite time limit, it is clear that in many cases (e.g. liquid water a few degrees Celsius below 0) the free energy does not depend on the time, when the time is larger than a very short microscopic time and it is in the range accessible by humans.

In the case of glasses we can start from an equilibrium configuration and define the time-dependent entropy $S(\tau)$ as the entropy of those configurations that can be reached in a time $\tau$. For $\tau$ much larger than  a microscopic time and smaller than the very large equilibration time we expect that $S(\tau)$ has a plateau and the value of this plateau (which  is not sharply  defined) is the entropy in one state (or valley), that can be used to compute the complexity using equation (\ref{EQ_CON}).

When the value of the equilibration time becomes infinite, i.e. at the Kauzmann transition
 (if it exists), the length of the plateau becomes infinite and its value becomes sharply defined. 

A similar game may be played in the framework of inherent structures. As noticed by Stillinger \cite{Stillinger}, the existence of many configurations that differ one from an other by a finite number of local moves leads to a complexity that never goes to zero at any non-zero temperature, just for the same reasons that forbid the standard entropy to go to zero at non-zero temperature.  In this case one should introduce a modified definition of  inherent structures in such a way as to avoid considering structures that are too similar. This can be done in many different ways.

In one approach one defines a minimum of order $k$ of the energy (e.g. $\sum_{i<j} v(r_i-r_j)$) as a local minimum
such that there are no configurations of lower energy that can be obtained by moving $k$ particles \cite{MeAspen}. The minima of order 
$0$ are 
(in a first approximation)
 equal to the standard inherent structure (e.g. local minima of the Hamiltonian) while minima of infinite order are global minima (the definition of minima that are stable with respect to the movement of $k$ particle is analogous  to the $k$-spin-flip-stable minima in spin glasses.
The number of minima of order $k$ is given by 
\be
\cN(k,V) \approx \exp(V \Sigma_k)
\ee
In the correct definition of the complexity one should assume that  $\Sigma_k$, as a function of $k$ has a plateau in $k$ in some region of $k<k_c$, where $k_c\gg 1$ is the minimum number of particles involved in a movement from one valley to an other valley (also $k_c$ should diverge at the Kauzmann transition). The value of the complexity is $\Sigma_k$ in the plateaux region. The {\em obvious} fact that in short range models 
\be
\lim_{k \to \infty} \Sigma_k = 0\, ,
\ee
cannot be used to argue against the existence of a plateau in $\Sigma_k$. Also in this case  the length of the plateau should go to infinity at the transition.
In the same spirit we could also define a $\Delta$ stable minimum, i.e. a local minimum that is separated by lower minima by an energy  barrier higher that  $\Delta$.

Although these last definitions (using the inherent structures) are not computationally handy, they provide an other consistent definition of the complexity. On the other hand the computation of the entropy of the valley can be easily implemented numerically.

In conclusions we have just shown that it is possible to give a precise, but 
{\em fuzzy} definition of complexity, that does not have shortcomings or inconsistencies. 
Notice that the necessity for this fuzzy definition has nothing to do with the replica approach: it is entirely due to the physical nature of the glassy metastability as discussed in Sect.\ref{sect:gmh}.

\section{The replica approach to structural glasses: general formalism} 

We shall explain here how the replica approach described in the previous section can be applied in practice to the first-principle study of some classes of structural glasses.
We 
keep here to the case of simple glass formers consisting of $N$ particles interacting by a pair potential $v(r)$ in a 
space of dimension $d$.
Explicit computations in high density fluids are rather involved already in the liquid phase. How can we hope to perform such a computation in the glassy phase? The approach based on the replica approach allows us to reduce the properties of the system in the low temperature phase to those of a slightly more complex, molecular, system in its liquid phase.

\begin{figure}
\centerline{\hbox{
\epsfig{figure=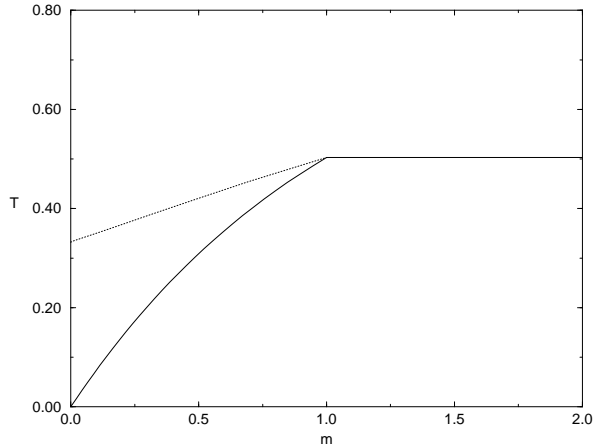,width=7 cm,angle=-90}
}}
\caption{ Typical phase diagram of a glass, replicated $m$ times,
in the limit of a  small inter-replica attractive potential,  in the plane $m$ - $T$ (temperature).
Above the full line the replicated fluid system is 
in its liquid phase, below this line it is a glass. 
 The dotted line is a first order transition 
line between a 
liquid which is a molecular fluid, where each molecule is a bound state of $m$ atoms of different color,
below the dotted line, and a liquid  where
the replicas don't make bound states,  above the line.
}
\label{fig13}
\end{figure}

Following the general approach explained in the previous section, the crucial point consists in studying the thermodynamics of a system made of $m$ replicas of the original  system, where we add an inter-replica attractive potential in such a way that the overlap between the different replicas is fixed to a large value\cite{MP,MePa1}. 
It will be useful to think of the $m$ replicas as `colors': in the replicated system, each atom is replicated $m$ times, and appears in $m$ distinct color states. What is the structure of this replicated fluid? If the inter-replica attractive potential would be infinitely strong, all the replicas of a given particle would be exactly at the same point. The total energy of a configuration wouls be $m$ times the energy of the unreplicated configuration occupying the same points in space. With a finite but large enough attractive potential the replicated system builds up a fluid of molecules, where each molecule is a bound state of $m$ atoms, each atom being one distinct color of the original atom.
If we focus on  the $m<1$ case, constraining $m$ replicas to be nearby decreases the energetic effects so is is natural to suppose that in the $m-T$ plane the glass transition point is shifted at lower temperature. The whole region $T>T_c(m,q)$ is in the high temperature phase and what happens there can be computed by generalizing the liquid approach (see Fig.\ref{fig13}). 
The free energy along the transition line at $m<1$ can be obtained from a high temperature computation in a correlated liquid. As the free energy in the glass phase is independent of $m$ (remember the behaviour of Fig.\ref{smqualit}), this allows to compute the full free energy in the glass phase at $m=1$. The idea will thus be to compute the free energy as function of $m$, for $m<1$, in the molecular fluid phase. From this free energy, by varying $m$, we will be able to locate the glass
transition line through the condition that $\Sigma=0$, and the value of the free energy at $m=1$ is equal to the value at the transition line.

\subsection{The partition function}
The usual partition function, used e.g. in the liquid phase, is 
\begin{equation}
Z_{1} \equiv {1 \over N!} \int \prod_{i=1}^N d x_i \  e^{-\beta H}\ ,
\end{equation}
where
\begin{equation}
H=\sum_{1 \le i<j \le N} v(x_i-x_j) \ .
\end{equation}
We wish to study the transition to the glass phase through the onset of an
off-diagonal correlation in replica space\cite{MP,MePa1}. We use $m$ replicas and  introduce
the Hamiltonian of the replicated system:
\begin{equation}
H_m=\sum_{1 \le i<j \le N} \sum_{a=1}^m v(x^a_i-x^a_j) +
\eps \sum_{i=1 N} \sum_{a<b =m}w(x^a_i-x^b_i) 
\end{equation}
where $w$ is an attractive interaction.  The precise form of $w$ is unimportant: it should 
be a short range attraction respecting the replica permutation symmetry, and its 
strength $\eps$
that will be sent to zero in the end.  For instance one could take
\begin{equation}
w(x)=- \ato  \frac{c^{2}}{x^{2+}c^{2}}\cto ^{6}
\end{equation}
with $c$ is of the order of 0.2 times the typical intermolecular distance. A positive value 
of $\eps$ forces $m$ particles, one from each color, to build a molecular bound state. If two systems stay 
in the same state, the expectation value of $w$ represents the self overlap and is very 
near to $1$.

The partition function of the replicated system is (leaving aside the trivial contribution from the kinetic energy):
\begin{equation}
Z_{m} \equiv {1 \over N!^m} \int \prod_{i=1}^N \prod_{a=1}^m d x^a_i \  
e^{-\beta H_m}
\end{equation}

\subsection{Molecular bound states}
At low enough temperature, we expect the following behaviour: If we prepare the system starting from large $\eps$,  so that we build the
molecular bound states of $m$ atoms (all replicas fall into the same glass state), and then decrease $\eps$,  the bound state will still exist also when $\eps\to 0$. This is a way to detect
the existence of a glass phase. In some sense this procedure is the analog for glasses of the usual symmetry breaking procedure which allows to detect a phase transition with a spontaneously broken symmetry. In an Ising ferromagnet in zero external field, we can detect the transition by measuring the magnetization in the presence of a small positive field, and letting the field go to $0^+$. If the result is non-zero, it is equal to the spontaneous magnetization. The same philosophy is used with our replica bound states. We form the bound states using the attractive potential at a finite $\eps$, and then see if the bound states still survive in the $\eps\to 0^+$ limit. If they do, the system is in a glass phase.

Thermal fluctuations are relatively small throughout the solid phase (one can see this from the Lindeman criterion) and 
diffusion is very small, so one can identify the molecules and relabel all the particles in the various replicas in such a 
way that the particle $j$ in replica $a$ always stays close to particle $j$ in replica $b$.  All the other relabelings 
are equivalent to this one, producing a global factor $N!^{m-1}$ in the partition function.

We therefore need to study a system of molecules, each of them consisting of $m$ atoms (one atom from each replica).  
It 
is natural to write the partition function in terms of the variables $r_i$ that describe the centers of masses of the 
molecules, and the relative coordinates $ u_i^a$, with $x_i^a=r_i+u_i^a$ and $\sum_a u_i^a=0$:

\bea \nn
Z_{m}&=& {1 \over N!} \int \prod_{i=1}^N dr_i  
 \prod_{i=1}^N \prod_{a=1}^m
d u_i^a \prod_{i=1}^N \((m^3 \delta(\sum_a u_i^a)\)) \\
& & \exp\((-\beta \sum_{i<j,a} v(r_i-r_j+u^a_i-u^a_j) - \beta
\sum_i \sum_{a,b}W(u^a_i-u^b_i) \))
\label{Zzu}
\eea

\subsection{ The small cage expansion}
In order to transform these ideas into a tool for doing explicit computations of the 
thermodynamic properties of a glass \cite{MePa1} we have to use an explicit method for estimating $Z_m$. This will give access to the 
free energy as function of the temperature and $m$, which gives the complexity by the Legendre transformation with respect to $m$.  As is usually the case, in the liquid 
phase exact analytic computations are not possible and we have to do some approximations.  
In this section we shall use the fact that the thermal fluctuations of the particles in 
the glass are small at low enough temperature: the size of the `cage' seen by each 
particle is therefore small, allowing for a systematic expansion.  What we will be 
describing here are the thermal fluctuations around the minimum of the potential of each 
particle, in the spirit of the Einstein model describing the vibrations of a crystal.
 
We start from the replicated partition function $Z_{m}$ described in molecular 
coordinates in (\ref{Zzu}). Assuming that the relative coordinates $u^a_i$
are small, we can expand $w$ to leading order and write:
\bea \nn
Z_{m}(\alpha)&=& {1 \over N!} \int \prod_{i=1}^N dr_i  \prod_{i=1}^N 
\prod_{a=1}^m
d u_i^a \prod_{i=1}^N \((m^3 \delta(\sum_a u_i^a)\)) \\
& & \exp\((-\beta \sum_{i<j,a} v(r_i-r_j+u^a_i-u^a_j) -
{1 \over 4 \alpha} \sum_i
\sum_{a,b} (u^a_i-u^b_i)^2 \))
\label{Zzu_exp}
\eea
In the end we are interested in the limit $\eps\equiv(1/\alpha) \to 0$.  We would like first to define the size $A$ of 
the molecular bound state, that is also a measure of the size of the cage seen by each atom in the glass, by:
\begin{equation}
{\partial \log Z_{m} \over \partial (1/\alpha)} \equiv {m (1-m) \over 2} d N A
=-{1 \over 4} \sum_i \sum_{a,b} \la (u_i^a-u_i^b)^2 \ra
\label{Adef}
\end{equation}
($d$ is the dimension that we have taken equal to 3 and $N$ is the number of particles).  We Legendre transform the 
free 
energy $\phi(m,\alpha)=-(T/m) \log Z_{m}$, considered as a function of $1/\alpha$, by introducing the thermodynamic potential per particle $\psi(m,A)$:
\begin{equation}
 \psi(m,A)= \phi(m,\alpha)+T d  { (1-m) \over 2} {A \over \alpha}
\end{equation}
What we want to see is whether there exists a minimum of $\psi$ at a finite value of $A$.

At low temperatures, this minimum should be at small $A$, and so we shall seek an expansion of $\psi$ in powers of 
$A$.  
It turns out that it can be found by an expansion of $\phi$ in powers of $\alpha$, used as an intermediate bookkeeping procedure
in order to generate the small $A$  expansion.  

This may look confusing since we are eventually going to send 
$\alpha$ to $\infty$.  
However this method is nothing but a usual low temperature expansion in the presence of an infinitesimal breaking 
field.  
For instance if one wants to compute the low temperature expansion of the magnetization in a $d$-dimensional Ising 
model 
in an infinitesimal positive magnetic field $h$, the main point is that the magnetisation is close to one.  One can 
organise the expansion by studying first the case of a large magnetic field, performing the expansion in powers of 
$\exp(-2h)$, and in the end letting $h \to 0$.  A little thought shows that the intermediate -large $h$- expansion is 
just a bookkeeping device to keep the leading terms in the low temperature expansion.  What we do here is exactly 
similar, the role of $h$ being played by $1 / \alpha$.

\subsubsection{Zeroth order term}
We use the equivalent form:
\begin{equation}
Z_{m}(\alpha)= {1 \over N!} \int \prod_{i=1}^N \prod_{a=1}^m
(d^3 u_i^a)
\prod_i {d^3X_i \over \sqrt{2\pi \alpha \over m^2}^3}
\exp\((-\beta \sum_{i<j, a} v(x_i^a-x_j^a)-{m \over 2 \alpha} \sum_{i,a} 
(x_i^a-X_i)^2\)) \ .
\end{equation}
In the limit $\alpha \to 0$, the identity 
\begin{equation}
\exp\((- {m \over 2 \alpha} (x_i^a-X_i)^2\)) \simeq \(({2 \pi \alpha \over 
m}\))^{d/2}
\delta^3 (x_i^a-X_i)
\end{equation}
implies that:
\begin{equation}
Z_{m}^0(\alpha) = \(({2 \pi \alpha \over m}\))^{3 N (m-1)/2} 
{1 \over N!} \int \prod_i dX_i \exp\((- \beta m \sum_{i<j} v(X_i-X_j) \)) \ .
\end{equation}
In this expression we recognise the integral over the $X_i$'s as the partition function $Z_{liq}(T^*)$ of the liquid at 
the effective temperature $T^*$, defined by
\begin{equation}
 T^*\equiv T/m \ .
 \end{equation} 
Therefore the free energy, at this leading order, can be written as:
\begin{equation}
{\beta \phi^0(m,\alpha)} = {3  (1-m) \over 2 m} \log {2 \pi \alpha \over m}
-{3  \over 2 m}
\log(m)-{1 \over mN} \log Z_{liq}(T^*)
\end{equation}
The result is intuitive: in the limits where the particles of different replicas stay at the same point, the 
Hamiltonian for $m$ replicas is just the usual one, multiplied by $m$.

\subsubsection{First order term}
\label{small_cage}
In order to expand to next order of the $\alpha^{-1}$ expansion, we start from  the representation 
(\ref{Zzu_exp})
and expand the interaction term to quadratic order in the relative coordinates:
\bea 
Z_{m}&=& \int \prod d^3r_i d^3u_i^a \prod_i \(( m^3 \delta(\sum_a u_i^a) \))
\exp\((-\beta m \sum_{i<j}
v(r_i-r_j)\))
\\ \nn
& & \exp\((-{\beta \over 2} \sum_{i<j} \sum_{a \mu \nu} (u_i^a-u_j^a)
(u_i^a-u_j^a)
{\partial^{2} v(r_i-r_j) \over \partial r^{2}}
-{1 \over 4 \alpha} \sum_{a,b} (u_i^a-u_i^b)^2\)) \ .
\label{Zzu_quad}
\eea
(where for simplicity we have not introduced the  indices $\mu$ and $\nu$, running from $1$ to $d$, 
that denote space directions).
Notice that in order to carry this step, we need to assume that the
interaction 
potential
 $v(r)$ is smooth enough, excluding hard cores.

After some  computations \cite{MePa1}, one finds that 
the free energy to first order is equal to:
\bea
{\beta \phi(m,\alpha)}&=& \\ &&{3  (m-1) \over 2 m} \log{1 \over \alpha} - \alpha
\beta 
 C +
{3 (1-m) \over 2 m} \log {2\pi \over m}-{3 \over 2 m } \log{m} -{1 \over mN}  
\log Z_{liq}(T^*)
\eea
where the constant $C$ is proportional to the expectation value
of  the Laplacian of the potential,
in the liquid phase at the temperature $T^*$: 
\begin{equation}
C \equiv  {1 \over 2} {1-m \over m^2} \sum_{j(\ne i)} \la \Delta v(z_i-z_j)\ra^*
\end{equation}
Differentiating the free energy  with respect to ${1 / \alpha}$ gives the equation for the size of the cage:
\begin{equation}
 {\beta }{\partial \phi \over \partial (1/\alpha)}= -{ (1-m) \over 2 m} d  
\alpha
+\alpha^2 \beta C=
 -{ (1-m) \over 2 } d  A
\end{equation}
Expanding this equation in perturbation theory in $A$ we have:
\begin{equation}
\alpha =m A -{ 2 \beta m^3 C \over 3 (m-1)} A^2
\end{equation}
The Legendre transform is then easily expanded to first order in $A$:
\bea\nn
 {\beta \psi(m,A) }&=& {\beta \phi}(m,\alpha) + 3 { (1-m) \over 2} {A \over 
\alpha}\\ &=& 
{3 (1-m) \over 2 m} \log(2 \pi A) - \beta m A C  +{3 (1-m) \over 2 m}
-{3 \over 2 m} \log m- {1 \over m N}  \log Z_{liq}(T^*)
\label{free_first_ord}
\eea

This very simple expression gives the free energy as a function of the number of replicas, $m$, and the cage size $A$.  
We need to study it at $m \le 1$, where we should maximise it with respect to $A$ and $m$.  
The fact that we seek a 
maximum when $m<1$ instead of the usual procedure of minimising the free energy is a well established fact of the 
replica method, appearing as soon as the number of replicas is less than $1$ \cite{mpv}.

As a function of $A$ , the thermodynamic potential $\psi$ has a maximum
at:
\begin{equation} 
A=A_{max} \equiv
{d(1-m) \over 2 \beta m^2 } {1 \over C} = {3 \over \beta} {1 \over \int d^3r 
g^*(r) \Delta v(r)}
\label{A_first_ord}
\end{equation}
where $g^*$ is the pair correlation of the liquid at the temperature $T^*$.  A study of the potential 
$\psi(m,A_{max})$, 
that equals $\phi(m)$, as a function of $m$ then allows to find all the thermodynamic properties that we seek, using 
the 
formulas of the previous section.  This step and the results will be explained below in sect.  \ref{results}, where we 
shall  compare the results to those of other approximations.

The systematic expansion of the thermodynamic potential $\psi$ in powers of $A$ can be carried out easily to higher 
orders.  However the result involves some more detailed properties of the liquid at the effective temperature $T^*$.  
For instance at second order one needs to know not only the free energy and pair-correlation of the liquid at 
temperature $T^*$, but also the three points correlation.  One can also obtain a partial resummation of the small cage expansion described above by integrating exactly over the 
relative vibration modes of the molecules.

\subsection{Harmonic resummation}
One can go beyond the first order small cage expansion by using the following method.
 Keeping only the term quadratic in $u$ in (\ref{Zzu_quad}) (harmonic vibrations of the molecules), and integrating
over
these vibration modes, one gets the "harmonic resummation"
approximation where the partition function is given by:
\be
Z_m= Z_m^0  \int dr
\exp\((-\beta m H(r) -{m-1 \over 2} Tr \log  M \)) \ .
\label{Zharmo}
\ee
Here $ Z_m^0 ={m^{Nd/2} \sqrt{2 \pi T}^{N d (m-1)} / N!}$, and
the matrix $M$, of dimension $dN \times dN$, is given by:
\be
M_{(i \mu) (j \nu)}= {\partial^2 H(r) \over \partial r_i^\mu \partial r_j ^\nu}
= \delta_{ij} \sum_k v_{\mu\nu}(r_i-r_k)-  
v_{\mu\nu}(r_i-r_j)
\ee
and $v_{\mu\nu}(r) =\partial^2 v /\partial r_\mu \partial r_\nu$ 
(the indices $\mu$ and $\nu$ denote space directions).
Now we are back to a real problem of liquid theory, since we have only $d$ degrees of freedom
per molecule (the center of mass coordinates), and the number of clones, $m$, appears
as a parameter in (\ref{Zharmo}). In particular there is no problem of principle to obtain results for arbitrary values of $m$ from this expression.

We have thus found an effective Hamiltonian for the centers of masses $r_i$ of 
the 
molecules, which basically looks like the original problem at the effective
 temperature $T^*=T/m$, complicated by the contribution of 
vibration modes which give the `Trace Log' term. 
We expect that this  should be a rather good approximation for the
glass phase. Unfortunately, even within this approximation,
it seems impossible to compute the partition function exactly. 
We can proceed by using a 'quenched approximation', i.e.
 neglecting the feedback of
vibration modes onto the centers of masses. This approximation becomes
 exact close to the Kauzmann temperature where $m \to 1$. The free energy is 
then:
\be
 {\beta \phi(m,T)} = -{d \over 2 m} \log(m)- { d (m-1) \over 2 m } \log(2 \pi)
-{1 \over m N} \log Z(T^*) +{m-1 \over 2 m} 
\la Tr \log \((\beta M  \)) \ra^*
\ee
which involves again the free energy and correlations of the liquid at the
temperature $T^*$.
Computing the spectrum of $M$ is an interesting problem of random matrix
theory, 
in
a subtle case where the matrix elements are correlated. Some efforts have been 
devoted
to such computations in the liquid phase where the eigenmodes are called 
instantaneous
 normal
modes \cite{INM1}. It might be possible to extend these approaches to our
case. A simple high density approximation, detailed in 
\cite{MePa1}, leads to the following expression for 
the replicated free energy:
\bea
 {\beta \phi(m,T)} &=& -{d \over 2 m} \log(m)- { d (m-1) \over 2 m } \log(2 \pi)
-{1 \over m N} \log Z(T^*) + {d (m-1) \over 2 m } \log (\beta r_0)
\\
\nn
&+&
{(m-1) \over 2 m } \intk  \(( L_3 \(({a(k)+{d-1 \over d} b(k) \over r_0}\))
+
{(d-1)} L_3 \(({a(k)-{1 \over d} b(k) \over r_0}\)) \)) \\ 
&-&
{(m-1) \over 4 m } \int d^dr g(r) \sum_{\mu\nu} {v_{\mu\nu}(r)^2 \over r_0^2}
\label{chain}
\eea
where the function $L_3$ is defined as:
\be
L_3(x)=\log(1-x)+x+{x^2/2}\ ,
\ee
and the Fourier transformed functions $a$ and $b$  are defined
from the pair correlation $g^*(r)$ by:
\be
\int d^d r \  g^*(r) v_{\mu\nu}(r) e^{ikr} \equiv \delta_{\mu\nu} \ a(k) +
\(( {k_\mu k_\nu \over k^2} -{1 \over d} \delta_{\mu \nu}\)) b(k) \ .
\label{defab}
\ee
We can thus compute the replicated free energy $F_m$ only from the knowledge
of the free energy and the pair correlation of the liquid
at the effective temperature $T^*$.
The results will be discussed in section \ref{results}.

\section {The replica approach to structural glasses: some results}
\label{results}
\subsection{Soft spheres}
\begin{figure}
\centerline{\hbox{
\includegraphics[width=7cm,angle=-90]{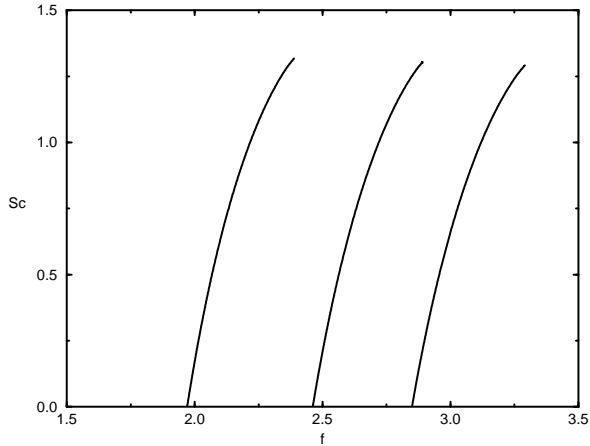}
}}
\caption{The configurational entropy $\Sigma(\beta,f)$ versus the free energy,
computed within the harmonic resummation, 
at temperatures
$T=1/\beta=0.,.05,.1$ (from left to right). }
\label{fig_Sc}
\end{figure}

We have studied \cite{MePa1} the case of soft spheres in three dimensions interacting
through a potential $v(r)=1/r^{12}$. We work for instance at unit density, since
the only relevant parameter is the usual combination $\Gamma=\rho T^{-1/4}$.
Fig. \ref{fig_Sc} shows the complexity versus free energy at various temperatures
computed with the harmonic approximation. We see that the main effect of changing the temperature is to shift 
the free energy.  This indicates that the main effect of temperature is to add a 
constant ($\approx 3/2 k T$) in the energies of all amorphous packings.  This correspond to the case where 
 the vibration spectrum is approximatively state independent.
Fig. \ref{fig_Sc} shows the internal energy and specific heat at various temperatures
computed with the three approximations explained above. The specific heat has a downward jump at the glass temperature,
from a value which is the liquid value to a value close to $1.5$ which is the Dulong Petit law (remember that we have left aside the
contribution from the kinetic energy), a very reasonable value for a solid within the Einstein model of vibrations in the classical limit. 
 Notice that it 
was not obvious at all a priori that we would be able to get such a result from our computations.  The fact of finding 
the Dulong-Petit law (from computations done in a molecular liquid!)
 is an indication that our whole scheme of computation gives reasonable results for a solid phase.
\begin{figure}
\includegraphics[width=.3\textwidth,angle=-90]{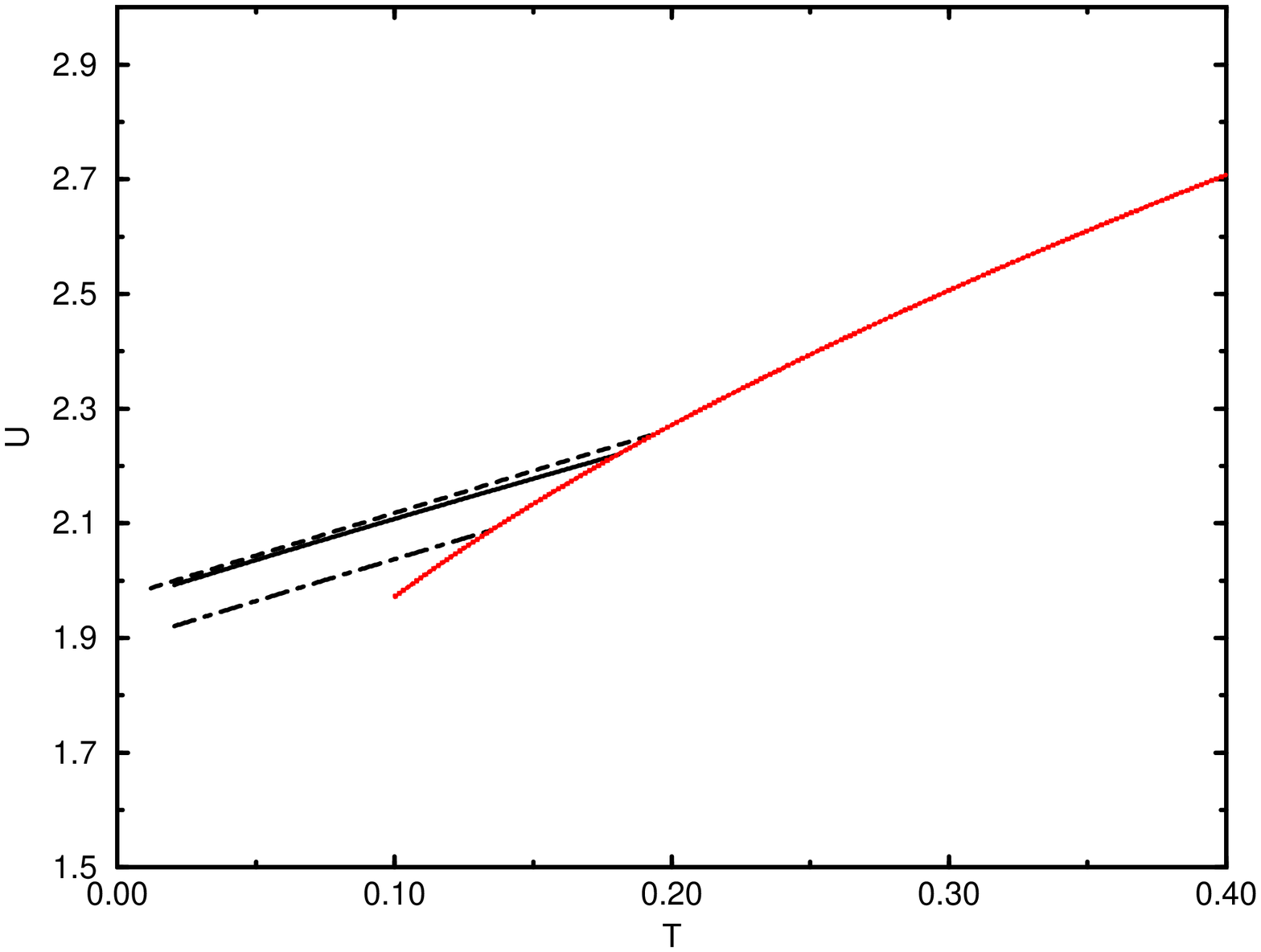}
\includegraphics[width=.3\textwidth,angle=-90]{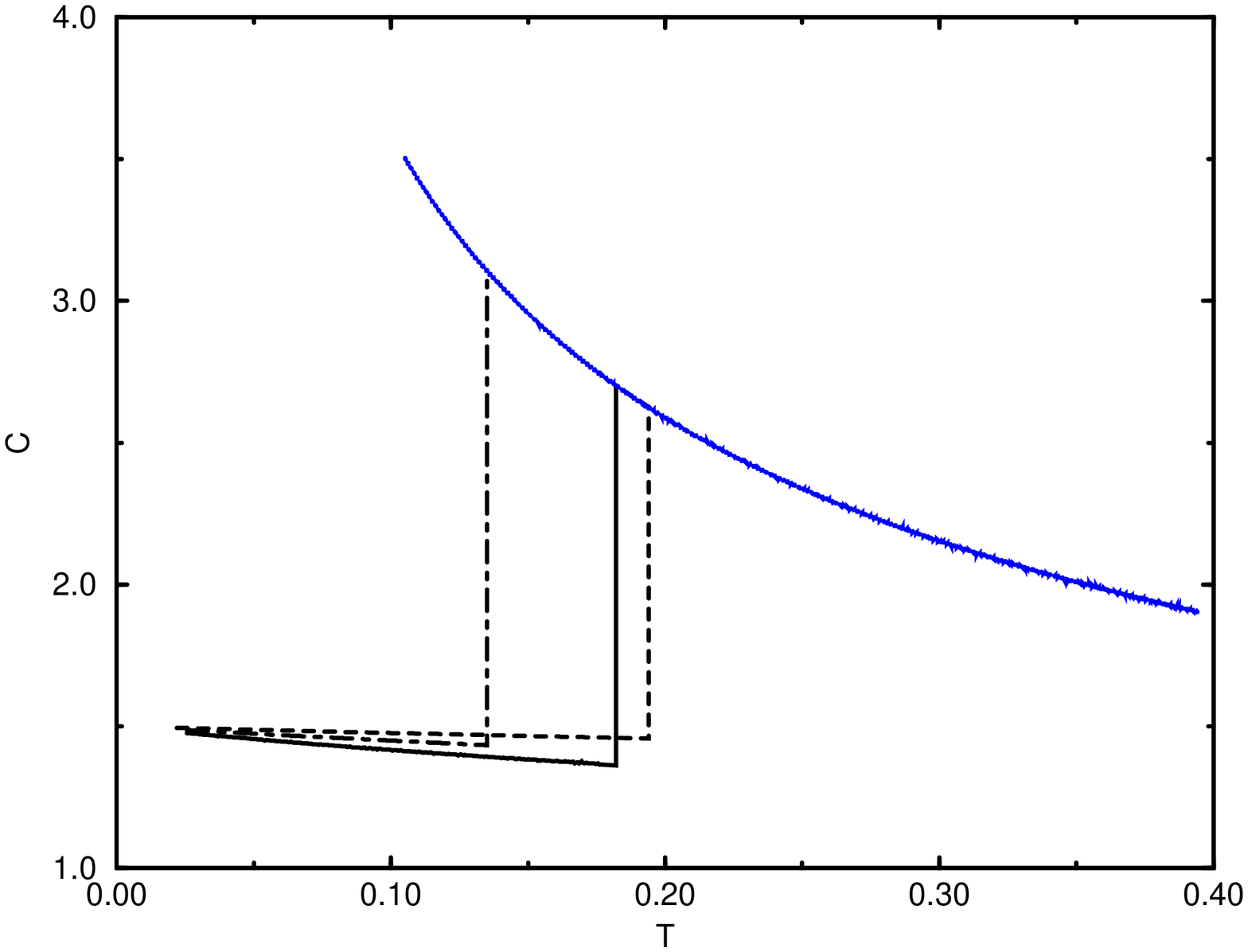}

\caption{The internal energy (left) and the specific heat (right) versus the temperature, for a system of soft spheres 
with unit density. These analytical results have been
computed in an expansion to first 
order (dashed-dotted line) and second order (full line) in the cage size $A$,
and in the harmonic resummation (dashed line). In the right panel the dotted line is the specific
heat of the liquid.}
\label{fig_C}
\end{figure}

\subsection{Binary mixtures of soft spheres}
The above results are physically very sound, but they
could not be directly compared to simulations because of the fact that soft sphere tend to crystallize too easily.
In order to be able to have a quantitative comparison with numerical data, we have extended 
the above approach to binary mixtures glass formers, where an appropriate choice of the interaction 
parameters
strongly inhibits crystallization. 

For soft spheres, we have studied the mixture of spheres introduced in
\cite{BAPA}\cite{HANSEN}, where the potentials are given by:
\be
V^{\eps\epp}(r)= \left ( \frac{\sigma_{\eps \epp}}{r} \right )^{12},
\ee
where
\be
\hspace{.3in} \frac{\sigma_{++}}{\sigma_{--}}=1.2, \hspace{.2in} 
\sigma_{+-}=\frac{\sigma_{++}+\sigma_{--}}{2} \ .
\label{pot}
\ee
The concentration is taken as $c_+=1/2$, and
the choice of the ratio $R\equiv\sigma_{++}/\sigma_{--}=1.2$ is known to strongly inhibit
crystallization.  
We also make the usual choice of considering particles with
average diameter 1 by setting
\begin{equation}
\frac{(\sigma_{++})^3+2(\sigma_{+-})^3+(\sigma_{--})^3}{4}=1.
\end{equation} 

All thermodynamic quantities depend on the density $\rho=N/V$ and the temperature $T$ only through the parameter $\Gamma \equiv \rho T^{-1/4}$. For $\Gamma$ larger than $\Gamma_D = 1.45$ (corresponding to lower temperatures) the dynamics becomes very slow and the autocorrelation time is very large.  Hence the system enters the 'aging' regime, where violations of the equilibrium fluctuation-dissipation theorem are observed \cite{FDT}.  This value of $\Gamma_D$ is supposed to correspond to the mode coupling transition below which the relaxation is dominated by
activated
processes \cite{HANSEN,BCKM}. 

\begin{figure}
\includegraphics[width=.35\textwidth,angle=-90]{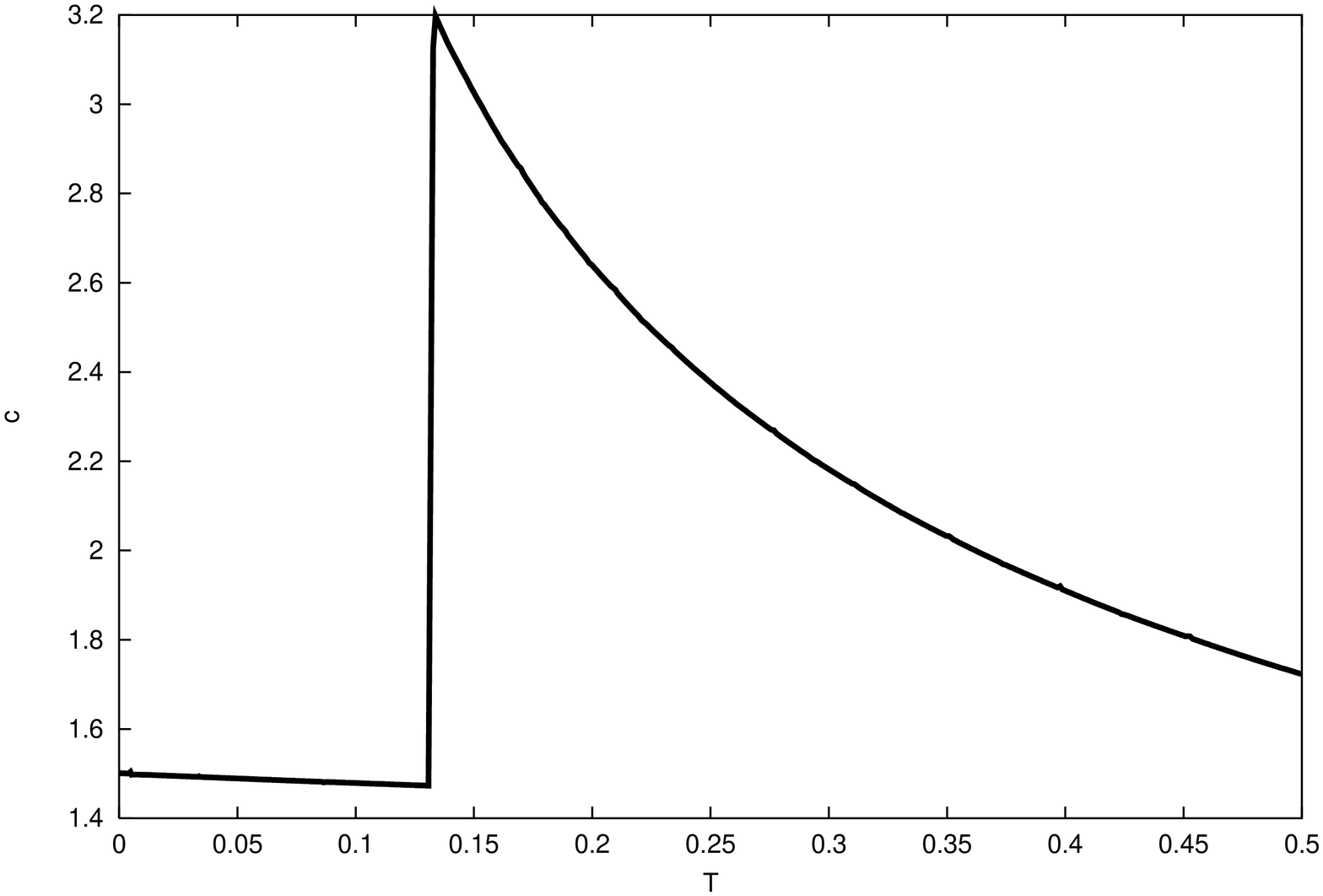}
\includegraphics[width=.35\textwidth,angle=-90]{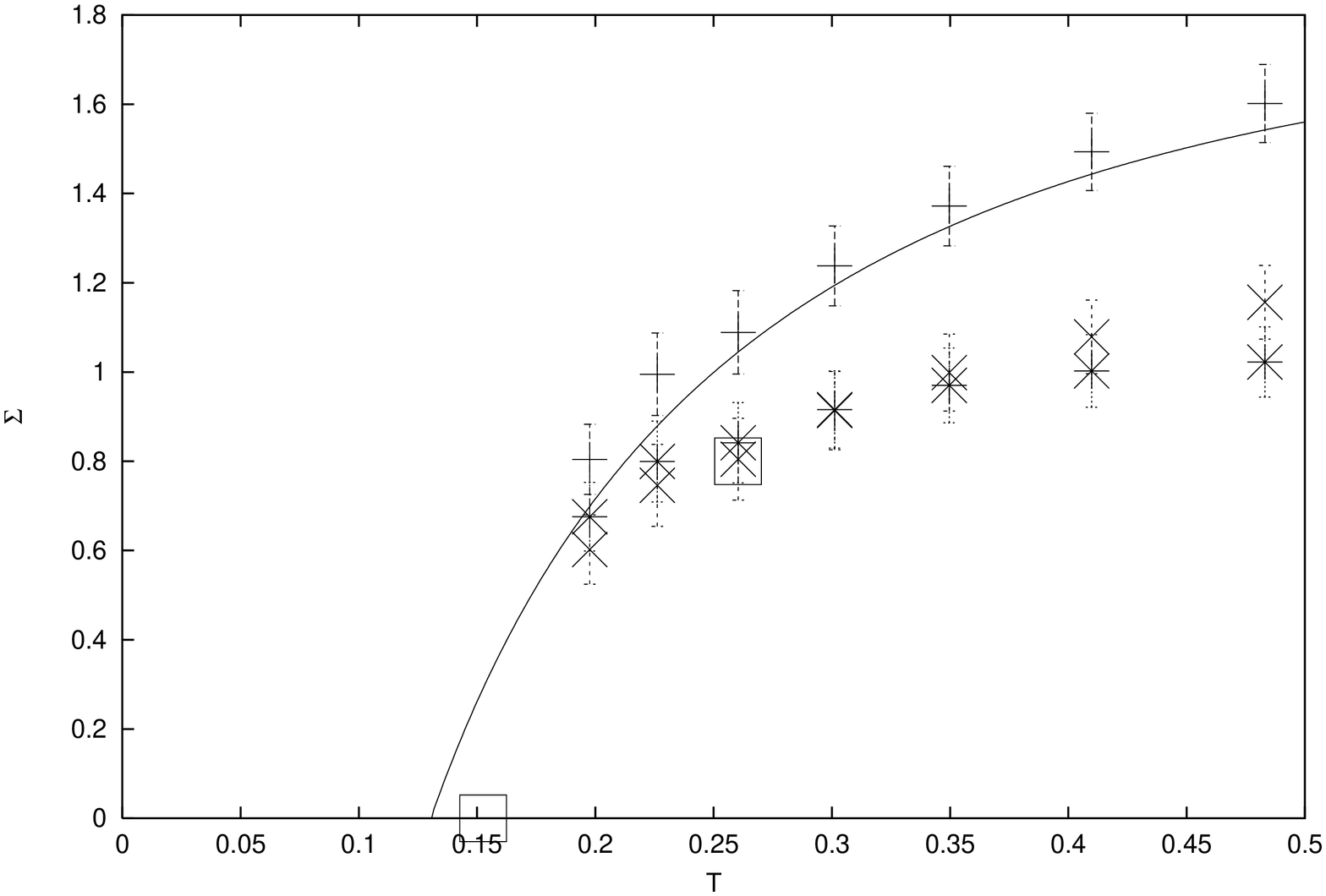}
\caption{ Results for binary mixtures of soft spheres. The top curve gives the specific heat,
the bottom curve gives the complexity. The full curve is the analytic prediction form the replica method. The points are the numerical estimates of the complexity obtained  with four different numerical methods: $s_{liq}-s^{(a)}_{sol}$ (+),
$s_{liq}-s^{(b)}_{sol}$ ($\times$),  a direct zero temperature Monte Carlo measurement($\ast$), and a study of the system coupled to a reference configuration ($\Box$).}\label{ssdat}
\end{figure}

The behaviour of the internal energy and the specific heat (shown in Fig.\ref{ssdat}) computed from the replica approach \cite{BIN} are very similar to those found in the pure soft sphere case. The equilibrium complexity $\Sigma(\beta,f^*(\beta))$ can also be computed, and we have compared it to some numerical estimates found as follows \cite{BIN}. The complexity is estimated from
 \be 
\Sigma(\beta)=\frac{1}{N} \left [ S_{liq} (\beta) - S_{sol}(\beta) \right ]\ .  
\ee 
The liquid entropy at temperature $T$ has been measured by numerical integration of the energy from infinite temperature down to $T$. Careful Monte Carlo simulations can be performed at high enough $T$, so that the thermalization can be controlled, and the data can be extrapolated with a power law towards lower temperatures.

As for the solid entropy associated to vibrations within one valley, we have estimated it numerically from
\be
{S_{sol}(\beta) \over N}= {d \over 2} (1+\log(2 \pi)) - {1\over 2 \: N}
\left \langle  \Tr \log (\beta {\cal M} ) \right \rangle \ ,
\ee
by diagonalizing the `instantaneous' Hessian $M$  and by averaging over different 
configurations. This is 
a  subtle task, as there always exists a non zero 
number of 
negative eigenvalues, which decreases as $\exp(-C/T)$ at low temperatures 
\cite{INM1} 
and is expected to be negligible below the Mode Coupling 
temperature. 
We have adopted the following
two measurements; the difference of their results gives an idea of the uncertainties associated with this procedure.
 a) One includes in the computation of
$\Tr \log (\beta {\cal M} )$ only the ${\cal N}_{pos}$ positive eigenvalues. b)
One includes all eigenvalues, but one takes the absolute 
values of 
the negative ones: 
\begin{equation}
{S^{(a)}_{sol} \over N} ={d \over 2} \left [ (1+\log({2 \pi \over \beta})) - 
\langle \frac{1}{{\cal N}_{pos}} \sum_{i=1}^{{\cal N}_{pos}} \log {\lambda_i} 
\rangle
\right ] \\
{S^{(b)}_{sol} \over N} ={d \over 2} \left [ (1+\log({2 \pi \over \beta})) - 
\langle \frac{1}{{ d N}} \sum_{i=1}^{d N} \log {|\lambda_i|}
\rangle \right ]. 
\end{equation}
Fig. \ref{ssdat} shows a good agreement between the analytical result of the replica method and the numerical simulations.

\subsection{Binary mixtures of Lennard-Jones particles}
 Another case of a more {\sl realistic} model for 
glasses is  the binary mixture of particles (80\% large particles, 20 \% smaller particles) interacting 
via a Lennard-Jones potential, introduced by Kob and Andersen \cite{KoAn}.  This Hamiltonian should 
mimic the behaviour of some metallic glasses and it is one of the best studied and simplest 
Hamiltonian which do not lead to crystallization at low temperature. The replica analysis performed in
\cite{LJ} again agree well with the numerical simulation results.

\begin{figure}
\includegraphics[width=.55\textwidth,angle=-0]{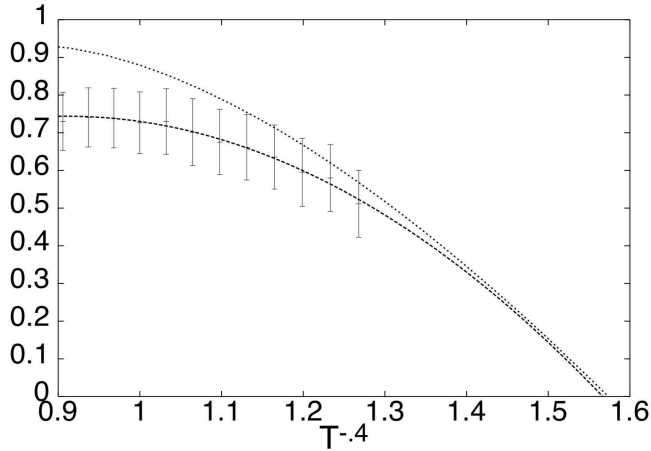}
\caption{ Analytical value of the complexity (upper line) compared with 
the numerical one ($+$ points), as functions of $\beta^{.4}$.}\label{DUE}
\end{figure}

Fig. (\ref{DUE}) shows in particular  the complexity as function of $T^{-.4}$.  The numerical result has been fitted with a 
polynomial of second degree in $T^{-.4}$.  The extrapolated configurational entropy becomes zero at 
a temperature $T_{c}=.31\pm.04$, where the error contains systematic effects due to the 
extrapolation (similar conclusions have been reached in ref.  \cite{KST}). This compares quite well with the result obtained from the
replica theory, $T_{c}\simeq.32$.

\subsection{Hard spheres}

Hard spheres have been used
as models for liquids, crystals, colloidal systems, granular, and
powders \cite{TOR}. Packings of hard spheres have a  wider interest: they
are related to important problems in information theory, such as
digitalization of signals, error correcting codes, and optimization
problems \cite{TWO,TWObis}. 
In particular, amorphous packings have attracted a lot of
interest as theoretical models for glasses, because for polydisperse
colloids and granular materials the crystalline state is not obtained
in experiments.

Hard spheres is the simplest theoretical model for glassy systems (although in three dimensions they do crystallize very fast in the monodisperse case). 

The replica approach can be used to study this problem, but  the simple small cage expansion of the previous sections does not work. Indeed we have seen that the first order correction in $A$ in the cage expansion is proportional to 
\be
 \sum_{j(\ne i)} \langle (\nabla V(z_i-z_j))^{2}\rangle \ .
 \ee
 The previous quantity is infinite for hard spheres, where the potential is discontinous and it is infinite for distances smaller than the diameter $D$. It easy to check that the divergence is not an artifact: if we approximate $V(x)$ by a continous potential (e.g. $V(x) =C(D-|r|)\theta(D-|r|)$ we obtain a divergence in the hard limit where $C \to \infty$.

This difficulty has a physical origin. Indeed   if one make some approximations valid at small $A$, but one waits to make the expansion in powers of $A$ at the very end \cite{FZ2005,FZ2006a,FZ2006b,FZ2008}, one obtains that the first corrections are proportional to $\sqrt{A}$, not to $A$. It is evident why the simple expansion in integer powers of $A$ is divergent. 

\begin{figure} \begin{center}    
      \includegraphics[width=.50\textwidth]{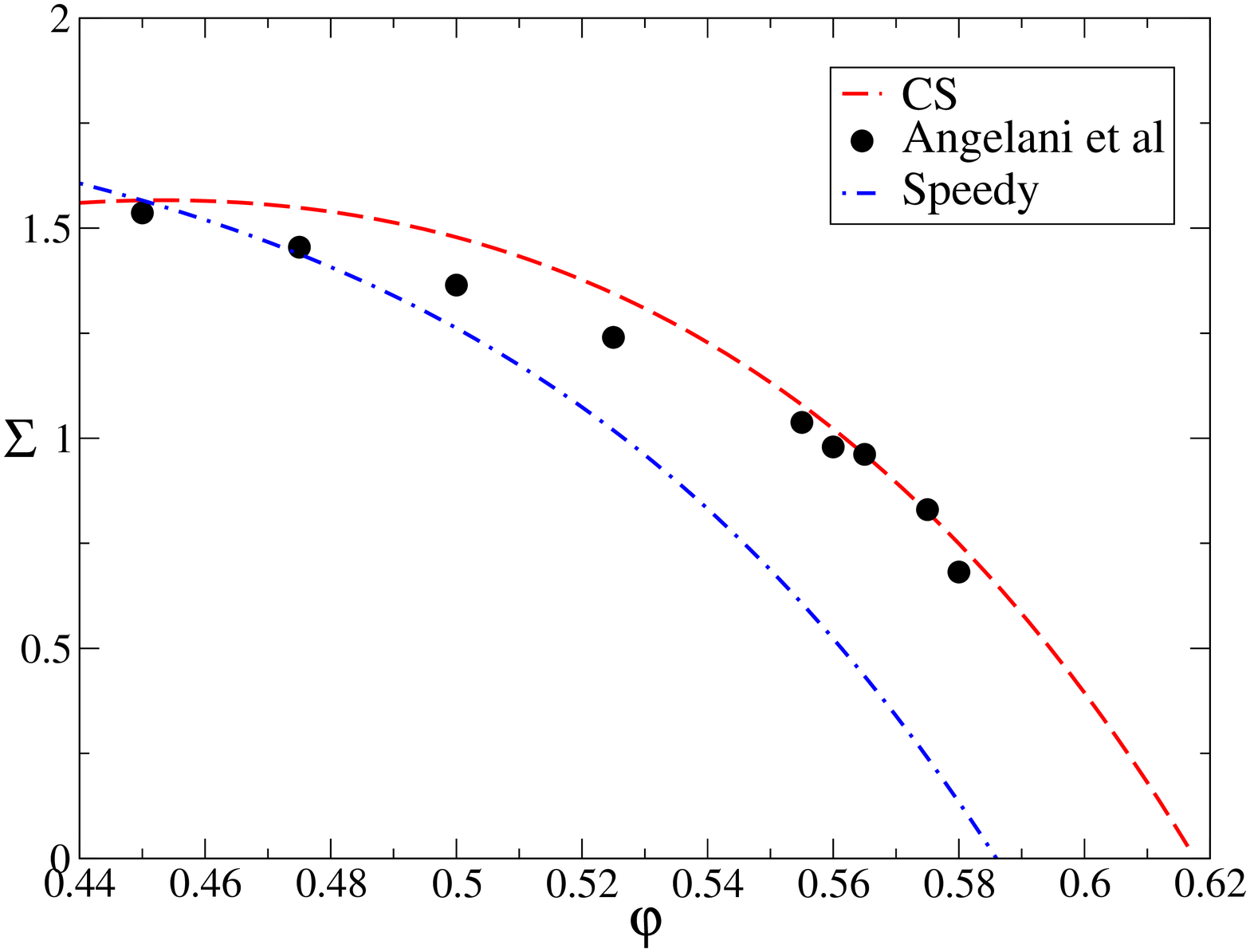}
      \end{center} \caption{ \label{F_CON}
The complexity of hard spheres in three dimensions, as a function of the packing fraction. The points are from numerical simulation\cite{AF},
the two curves are the results of the analytic replica approach, in which the {\em liquid} phase properties have been computed either with the Carnahan-Starling ("CS") approach\cite{CS}, or with the one of Speedy \cite{Speedy}.
}\end{figure}

Using only the first term in the new expansion one obtains remarkably good results\cite{FZ2005,FZ2006a,FZ2006b,FZ2008}. For example we see in fig. (\ref{F_CON}) the computation of the complexity for three dimensional hard spheres as function of the  packing fraction $\phi$, compared with the results coming from numerical simulations \cite{AF}.

This approach allows a detailed computations of many properties of hard spheres system, including  those of jammed packing \cite{FZ2008}, also for binary mixtures and four dimensional models. The results are usually in a remarkable good agreement with the numerical data. A very interesting case is the computation of detailed properties of binary mixtures like the number of contact among particles of equal and different radius \cite{BCPZ}.

\section{Conclusion}

Summarizing, we have described a simple method that is able to use standard methods from liquid theory in order
to derive properties of the glass phase,
putting in practice the old adage {\sl a glass is a frozen liquid}.  The method uses replicas of the original system, where, in a sense, each replica provides a polarizing field for the other replicas. This seems to be at the moment the best way to study problems, like glasses, which can be in many metastable states. Using this method, we have shown how  to compute with some 
reasonable approximations the thermodynamics and, with a little more effort, we can compute the equilibrium correlation functions. 
Within the equilibrium framework, we have implemented so far our general strategy using rather crude methods.  These 
methods can be systematically improved, and one could certainly perform a more careful study of the molecular liquid, using more sophisticated methods from liquid theory. 

\section*{Ackowledgments} The replica approach to structural glasses has been developed in
collaboration with many colleagues and friends. It is a great pleasure
to thank them.

\end{document}